\documentclass[traditabstract]{aa}

\usepackage{graphicx}
\usepackage{txfonts}
\usepackage{natbib}
\bibpunct{(}{)}{;}{a}{}{,} 

\usepackage{longtable}
\usepackage{pdflscape}
\usepackage{wasysym}
\usepackage{url}

\newcommand{\au}{{\sc au}}
\newcommand{\micron}{$\mu$m}

\newcommand{\etal}{et al.}

\newcommand{\Mearth}{M$_\oplus$}
\newcommand{\Rearth}{R$_\oplus$}

\newcommand{\hst}{\emph{HST}}
\newcommand{\spitzer}{\emph{Spitzer}}

\begin{document}

	\title{Near-infrared transmission spectrum of the warm-uranus GJ~3470b with the Wide Field Camera-3 on the \emph{Hubble Space Telescope}}

	\author{D.~Ehrenreich\inst{1}, X.~Bonfils\inst{2}, C.~Lovis\inst{1}, X.~Delfosse\inst{2}, T.~Forveille\inst{2}, M.~Mayor\inst{1}, V.~Neves\inst{2,3,4}, N.~C.~Santos\inst{3,4}, S.~Udry\inst{1}, \&~D.~S\'egransan\inst{1}} 

    \institute{
      Observatoire de l'Universit\'e de Gen\`eve, 51 chemin des Maillettes, 1290 Sauverny, Switzerland	
      \email{david.ehrenreich@unige.ch}
      \and
      Universit\'e Joseph Fourier-Grenoble~1, CNRS-INSU, Institut de Plan\'etologie et d'Astrophysique de Grenoble (IPAG), UMR~5274, BP~53, 38041 Grenoble, France
      \and
      Departemento de F\'\i sica e Astronomia, Faculdade de Ci\^encias, Universidade do Porto, rua do Campo Alegre, 4169-007 Porto, Portugal
      \and
      Centro de Astrof\'\i sica, Universidade do Porto, rua das Estrelas, 4150-762 Porto, Portugal}

    \date{}

   \abstract{The atmospheric composition of super-earths and neptunes is the object of intense observational and theoretical investigations. Transmission spectra recently obtained for such exoplanets are featureless in the near infrared. This flat signature is attributed to the presence of optically-thick clouds or translucent hazes. The planet \object{GJ~3470b} is a warm neptune (or uranus) detected in transit across a bright late-type star. The transit of this planet has already been observed in several band passes from the ground and space, allowing observers to draw an intriguing yet incomplete transmission spectrum of the planet atmospheric limb. In particular, published data in the visible suggest the existence of a Rayleigh scattering slope -- making GJ~3470b a unique case among the known neptunes, while data obtained beyond 2~\micron\ are consistent with a flat infrared spectrum. The unexplored near-infrared spectral region between 1~\micron\ and 2~\micron, is thus the key to understanding the atmospheric nature of GJ~3470b. Here, we report on the first space-borne spectrum of \object{GJ~3470}, obtained during one transit of the planet with the Wide Field Camera-3 (WFC3) on board the \emph{Hubble Space Telescope} (\hst), operated in stare mode. The spectrum covers the 1.1--1.7~\micron\ region with a resolution of $\sim 300$ ($\Delta\lambda \sim 4$~nm). We retrieve the transmission spectrum of GJ~3470b with a chromatic planet-to-star radius ratio precision of 0.09\% (about half a scale height) per 40~nm bins. At this precision, the spectrum appears featureless, in good agreement with ground-based and \emph{Spitzer} infrared data at longer wavelengths, pointing to a flat transmission spectrum from 1~\micron\ to 5~\micron. We present new simulations of possible theoretical transmission spectra for GJ~3470b, which allow us to show that the \hst/WFC3 observations rule out cloudless hydrogen-rich atmospheres ($>10\sigma$) as well as hydrogen-rich atmospheres with tholin haze ($>5\sigma$). Adding our near-infrared measurements to the full set of previously published data from 0.3~\micron\ to 5~\micron, we find that a cloudy, hydrogen-rich atmosphere can explain the full transmission spectrum: the tentative Rayleigh slope in the visible and the flat near-infrared spectrum can be both reproduced if the water volume mixing ratio is lower at the terminator than predicted by equilibrium thermochemistry models.}


    \titlerunning{\hst/WFC3 transmission spectrum of GJ~3470b}
    \authorrunning{Ehrenreich \etal}

    \maketitle

\section{Introduction}

\subsection{What is the nature of exoplanets with masses between super-earths and neptunes?}

Exoplanets with measured masses and sizes similar to our ice giants, Uranus and Neptune, are still rare among the bestiary of nearly 300 known and confirmed transiting exoplanets ($\sim30\%$ of all confirmed exoplanets). Prototypical examples are the warm neptunes \object{GJ~436b} \citep{Butler:2004ki} and \object{HAT-P-11b} \citep{Bakos:2010ec}. The epithet warm is used in reference to the equilibrium temperature of these objects, between 500~K and 1\,000~K, which is cooler than the typical temperatures of hot jupiters.

With the discovery of super-earths such as \object{GJ~1214b} \citep{Charbonneau:2009em} or \object{HD~97658b} \citep{Dragomir:2013ga,VanGrootel:2014je}, a mass continuum is observed from 1~\Mearth\ to 20--30~\Mearth. This low-mass continuum seems to stop at $\sim30$~\Mearth, with a gap extending up to the mass of Saturn that is attributed to the runaway accretion of gas in the core accretion model of planet formation \citep[e.g.][]{Mordasini:2009ff}.

Planets at the extremities of the apparent low-mass continuum should, however, bear significant differences in their structures and formation history. For once, telluric planets should be dominated by their solid cores while Neptune-like objects are dominated by their envelopes of gas (H/He) and volatiles \citep[water is generally considered the dominant volatile species, although see other possibilities in][]{Hu:2014gz}. Consequently, a transition is expected between these two different structures. This transition is not driven by the planet mass alone, as the detection of low-density super-earths, such as Kepler-11f \citep{Lissauer:2011el}, would tend to show.

One important feature in this core-to-envelope transition is the presence or absence of hydrogen and helium from the envelope \citep[e.g.,][]{MillerRicci:2010dz}. Observationally, this is a critical point because a H/He-rich atmosphere has a small mean molar mass $\mu$ and thus a high atmospheric scale height ($\propto \mu^{-1}$). In a clear atmosphere (cloud- or haze-free), a high atmospheric scale height favours the detection of absorption from volatile molecules (mainly water) in the near-infrared transmission spectrum of the planetary atmosphere, which can be retrieved during a planetary transit \citep{Seager:2000iy}. On the contrary, envelopes devoid of H/He but dominated by the volatile molecules in question, have scale heights typically an order of magnitude smaller than H/He-rich envelopes, and should present flatter transmission spectra.

Practically, this spectral discrimination between H/He-rich and volatile-rich planets is complicated by the presence of high-altitude clouds or hazes, which effectively hide the molecules in the lower atmosphere and produce near-infrared transmission spectra dominated by Mie scattering \citep{Ehrenreich:2012iq,Howe:2012et}, i.e. flat. While it should be theoretically possible to infer the presence of scattering from transmission spectra extending from the blue ($\sim 400$~nm) down to the infrared ($\sim 2~\mu$m), in practice it is difficult to collect such data. Broad-band spectra have nevertheless been obtained for the most studied exoplanets, the hot Jupiter HD~209458b \citep{Sing:2008ce,Deming:2013ge} and HD~189733b \citep{Pont:2013ch,Huitson:2012hk} using the STIS and WFC3 spectrographs on board the \emph{Hubble Space Telescope}. These spectra are clearly affected by scattering. The recent WFC3 transmission spectrum of HD~209458b by \citep{Deming:2013ge} shows a smaller-than-expected absorption of the water molecular band around 1.38~\micron. This is attributed to scattering by hazes, although in the case of HD~209458b the scattering slope increasing towards short wavelengths seems less steep than in the case of HD~189733b \citep{Deming:2013ge}.

The 1.38~\micron\ water band, known as the $\varphi$ band separating the $J$ and $H$ near-infrared windows in the Earth atmosphere, has been searched for in the transmission spectrum of GJ~1214b \citep{Charbonneau:2009em}, a low-density (1.9~g~cm$^{-3}$) low-mass (6.6~\Mearth) planet transiting a late M dwarf \citep{Berta:2012ff}. The \hst/WFC3 spectroscopic data of \citep{Berta:2012ff} show that the transmission spectrum of GJ~1214b is flat, within the $\sim1\,000$~ppm error bars (in $R_p/R_\star$), between 1.1~\micron\ and 1.7~\micron\ \citep[as confirmed by][]{Benneke:2013hu}. Additional ground-based \citep{Bean:2010ct,Bean:2011dc,Crossfield:2011fl,Croll:2011fv} and \emph{Spitzer} \citep{Desert:2011fy} measurements unveil a flat, featureless transmission spectrum from 0.6~\micron\ to 4.5~\micron, pointing at either a high-molecular-weight envelope or the presence of high-altitude haze masking the spectral features of the underlying atmospheric species in a hydrogen-rich atmosphere \citep{Rogers:2010fv,MillerRicci:2010dz,MillerRicciKempton:2012hn,Nettelmann:2011dh}. 

Recently, \citet{Kreidberg:2014du} have reported on a near-infrared transmission spectrum of GJ~1214b obtained from the combination of a dozen of transits observed with \hst/WFC3. Taking advantage of the co-addition of $\sim4\times$ more data and an increased observing efficiency compared to \citet{Berta:2012ff}, these authors have been able to extract a transmission spectrum with a resolution of about 70 at the exquisite precision of $\sim 30$~ppm per spectral bin. This data set has allowed \citet{Kreidberg:2014du} to rule out both a hydrogen-dominated atmosphere and a volatile-rich atmosphere dominated by water at a significant level, making it highly probable that clouds are responsible for the flatness of the transmission spectrum. In the case of the warm neptune GJ~436b, \citet{Knutson:2014hz} rule out a cloud-free hydrogen-dominated atmosphere based on \hst/WFC3 spectroscopy.

\subsection{The case of GJ~3470b}
In order to understand the class of warm exoplanets intermediate between super-earths and neptunes, and in the perspective of the rise of future instruments dedicated to probe exoplanetary atmospheres in this wavelength range (in particular the \emph{James Webb Space Telescope}), it is essential to obtain observational insights on new transiting members of this family of objects, and determine whether flat spectra are the rule or the exception. 

The planet GJ~3470b \citep{Bonfils:2012gb} is one such object, more massive than super-earths like GJ~1214b or HD~97658b, but lighter than a warm neptune like GJ~436b. The planet was discovered around a M1.5 dwarf during a velocimetric campaign with the HARPS spectrograph and detected in transit with the TRAPPIST, Euler, and NITES telescopes. A high-quality transit light-curve has been subsequently obtained with \emph{Spitzer} \citep[][the planetary parameters we will refer to in the following derive from the \spitzer\ data analysis]{Demory:2013hv}.

GJ~3470b is close to Uranus (\uranus) in terms of mass with $M_p = 13.9\pm1.5$~\Mearth\ (vs.\ $M_\textrm{\uranus} = 14.505$~\Mearth), yet it is significantly less dense ($\rho_p = 0.72\pm0.13$~g~cm$^{-3}$) than the Solar System ice giant ($\rho_\textrm{\uranus} = 1.27$~g~cm$^{-3}$), owing to a larger radius ($R_p = 4.83\pm0.22$~\Rearth\ vs. $R_\textrm{\uranus} = 3.968$~\Rearth). The planet is of course warmer than Uranus, with an equilibrium temperature of $T_p \sim 600$~K (for an albedo of 0.3) owing to its semi-major axis of $a_p = 0.035$~\au.

The planet has been recently observed in several optical to near-infrared photometric bands ($g'$, $R$, $I$, and $J$) with the 50~cm and 188~cm telescopes at Okayama Astrophysical Observatory, allowing \citet{Fukui:2013hq} to claim significant variations in the reconstructed broad-band transmission spectrum, and thus for (a hint for) a cloud-free atmosphere. This conclusion has, however, been challenged by \citet{Crossfield:2013eu}: these authors use the new MOSFIRE spectrograph installed at the Keck~{\sc i} telescope to produce a transmission spectrum of GJ~3470b between 2.09~\micron\ and 2.36~\micron. Combining these data with the broad-band measurements of \citet{Fukui:2013hq} at shorter wavelengths and \citet{Demory:2013hv} at 4.5~\micron, \citet{Crossfield:2013eu} conclude to a flat transmission spectrum, suggestive of the presence of high-altitude haze or (extremely) high metal-rich composition, much like \object{GJ~1214b}. 

\citet{Nascimbeni:2013db} announced two additional measurements of \object{GJ~3470} planet-to-star radius ratio, obtained from simultaneous transit observations in the near-ultraviolet and near-infrared with the Large Binocular Telescope (LBT). In contrast with the conclusion of \citet{Crossfield:2013eu}, the LBT data points are compatible with a steep slope of the transmission spectrum in the visible. \citet{Nascimbeni:2013db} attribute this slope to scattering by haze, basing their claim on the scaling of a theoretical transmission spectrum calculated for an atmosphere covered by a high-altitude haze of tholins \citep{Howe:2012et}. While such a model could produce a flat transmission spectrum in the infrared (for a given set of haze properties) and a steep rise in the optical, the scaled version of this model as presented by \citet{Nascimbeni:2013db}, seems unable to reconcile both data sets.

In this paper, we present the first space-borne measurement of \object{GJ~3470b} transmission spectrum. The \hst/WFC3 spectrum ranges from 1.1~\micron\ to 1.7~\micron, exploring the 1.38~\micron\ water band and providing a direct comparison with the work of \citet{Berta:2012ff} for \object{GJ~1214b}. This spectral region bridges the gap between the studies of \citet{Nascimbeni:2013db} and \citet{Fukui:2013hq} on one hand, and \citet{Crossfield:2013eu} and \citet{Demory:2013hv} on the other hand. It additionally overlap with the $J$-band measurement of \citet{Fukui:2013hq}, which anchors these authors claim for a transparent atmosphere.

We present the \hst/WFC3 observations in Sect.~\ref{sec:obs} and the data processing and analysis in Sect.~\ref{sec:data}. The broad-band near-infrared radius of \object{GJ~3470b} and the transmission spectrum obtained are detailed in Sect.~\ref{sec:results}. In Sect.~\ref{sec:discu}, we discuss our results with respect to those previously obtained and provide a theoretical interpretation based on dedicated calculations of transmission spectra for several atmospheric models.

\section{\hst/WFC3 observations}
\label{sec:obs}

We obtained director's discretionary time to observe one transit of \object{GJ~3470b} with \hst/WFC3 (programme GO/DD~13064) on 2013-Feb-07. The visit was split between 5 \hst\ orbits (one orbit lasts for 96~min). During each \hst\ orbit, the target star was visible for about 52~min; it was occulted by the Earth during the remaining time, yielding gaps in the light curve. The observation timing has been set so that all observations in the third \hst\ orbit would be acquired while GJ~3470b is fully transiting, i.e., between the second and third contact of the transit (GJ~3470b fully transits for 1.6~h). Observations during the first two \hst\ orbits have been acquired before the first contact and observations during the last two \hst\ orbits have been acquired after the fourth contact of the transit, providing an adequate baseline to fit the transit light curve. However, no data have been obtained during the transit egresses (between the first and second contacts or the third and fourth contacts).

\begin{figure*}[!ht]
\resizebox{\textwidth}{!}{\includegraphics[trim=0.5cm 1.5cm 1cm 7cm]{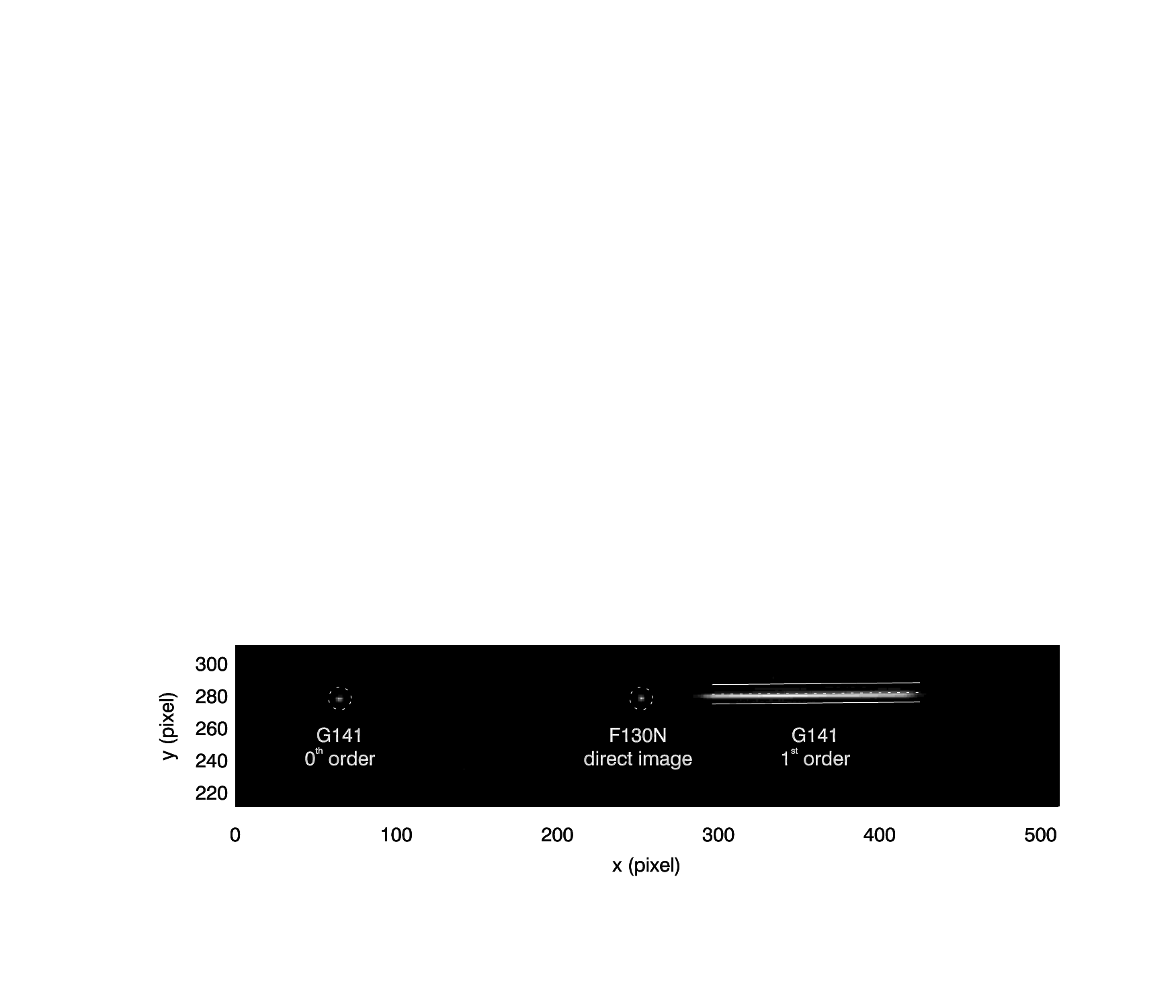}}
\caption{\label{fig:image} Example of a WFC3/G141 spectral image with the undispersed stellar image ($0^\mathrm{th}$ order and $1^\mathrm{st}$-order spectrum. The associated F130N direct image is superimposed. The spectral trace is shown by a dotted line that extends in the dispersion direction over a wavelength range 1.1--1.7~\micron. The plain lines above and below the trace show the width of the extraction region.}
\end{figure*}

The observations are grism slit-less spectroscopy. They consist in series of exposures taken with the G141 grism of WFC3 infrared channel (WFC3/IR) in stare mode. One example is shown in Fig.~\ref{fig:image}. The grism provides a spectral coverage of 1.1--1.7~\micron, including the 1.38~\micron\ band of water with a resolution of $\sim 300$. WFC3/IR exposures are obtained in \texttt{MULTIACCUM} mode, i.e. they are built from multiple non-destructive readouts of the Teledyne HgCdTe detector \citep{Dressel:2010tl}.

Because GJ~3470 is a bright source in the near-infrared ($J = 8.8$, $H = 8.2$), these WFC3 observations have a poor duty cycle, as already reported by \citet{Berta:2012ff} in the case of GJ~1214. Large overheads result from serial downloads of the WFC3 image buffer, during which new exposures cannot be started. The image buffer can only hold two full-frame images, therefore windowing of the detector is mandatory to minimise the number of buffer downloads. On the other hand, we wanted to keep the target image ($0^\mathrm{th}$ order) together with the first-order spectrum on the images. This was possible using a $512\times512$ subarray (\texttt{GRISM512}). The \texttt{RAPID} sampling sequence advised for bright targets was used to enhance the  observing efficiency, while the number of samplings $\texttt{NSAMP}$ was set to 3 to avoid saturation. These settings yield an effective exposure time of 2.559~s, which is $\sim70\%$ of the detector saturation time. Two batches of 32 exposures and one batch of 6--9 exposures, separated by 11~min serial buffer dumps could have been acquired during each \hst\ orbit. (The target acquisition prevented us from acquiring a third batch of exposure during the first \hst\ orbit.) In total, we obtained 350 exposures. 

For each orbit, the batches of spectroscopic G141 exposures are flanked by two direct images of the star obtained with the F130N narrow filter obtained with one sampling ($\texttt{NSAMP} = 1$) of the \texttt{RAPID} sampling sequence, yielding an integration time of 0.85~s.

The resulting observing efficiency of this programme is 6\%, to be compared with the 10\% obtained by \citet{Berta:2012ff} in the case of GJ~1214 -- a fainter target accounting for the slightly higher efficiency. We note that after this observing programme was designed, a new mode for WFC3/IR grism spectroscopy referred to as scanning mode has been made available, that allows much more efficient observations \citep{Deming:2013ge,Kreidberg:2014du,Knutson:2014hz}.

\section{Data treatment and analysis}
\label{sec:data}

The data analysis of these \hst/WFC3 data is similar in many respects to the treatment of GJ~1214 \hst/WFC3 by \citet{Berta:2012ff}. Our starting point for the analysis is the flat-fielded (FLT) images produced by \texttt{CALWF3}, the WFC3 pipeline. The analysis follows the steps described below.

\subsection{Extraction of the stellar spectra}

\subsubsection{Identification of cosmic ray hits}
\label{sec:cosmics}
\begin{figure}
\resizebox{\columnwidth}{!}{\includegraphics{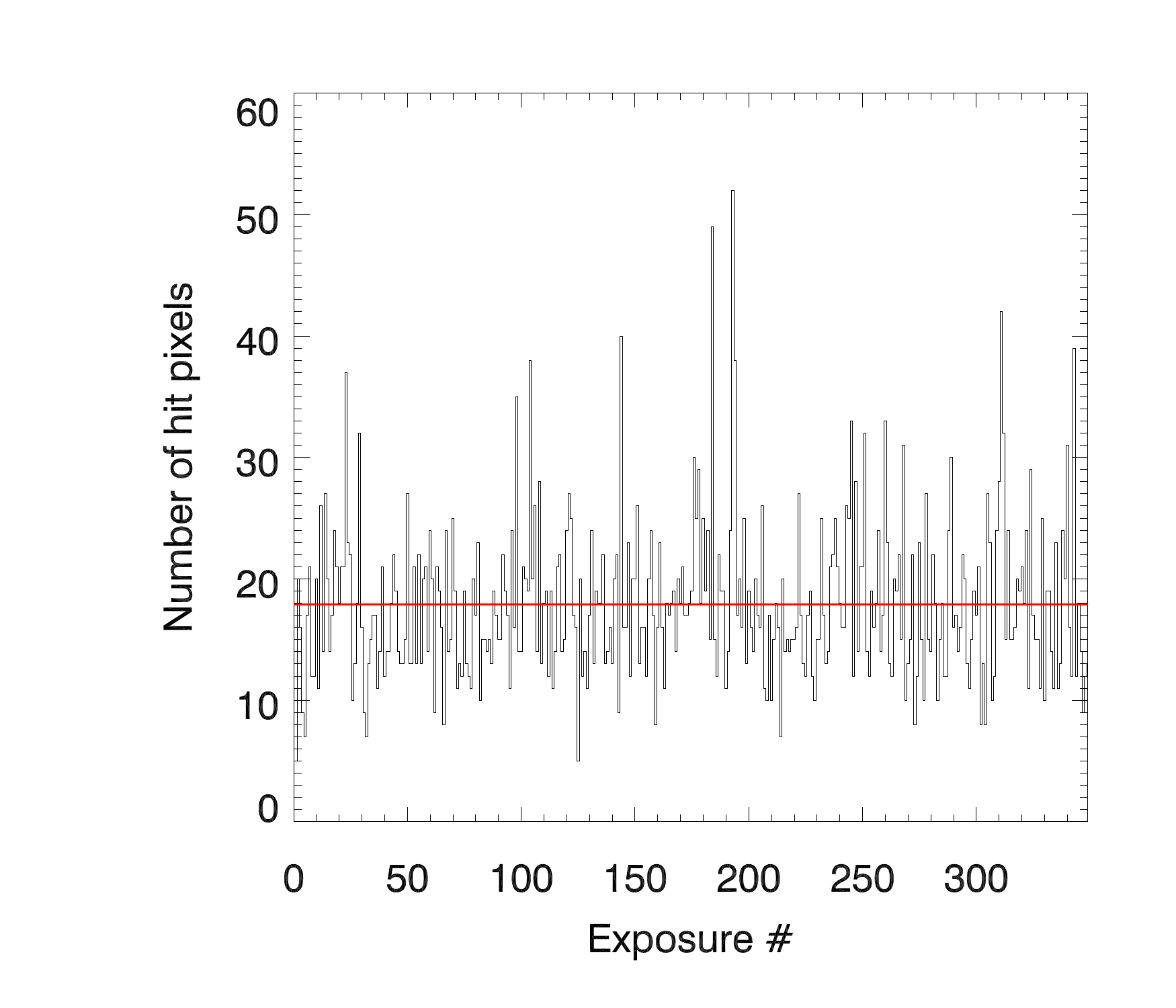}}
\caption{\label{fig:crhit}Number of pixels hit by cosmic rays as a function of the exposure number (i.e., time). The horizontal red line show the average value of 17.8 pixels throughout the whole sequence.}
\end{figure}

We fit the value of each pixel through our 350 images with a fourth-order polynomial of the frame number. The residuals of the fit are then compared to the standard deviation of the time series obtained for each pixel: at any time, a pixel with a value that is more than four times the standard deviation is considered to be hit by a cosmic ray. The pixel value is replaced by the value of the fit at this instant. The number of pixels hit by cosmic rays as a function of time is represented in Fig.~\ref{fig:crhit}. This method allowed us to identify an average of 17.8~pixels affected by cosmic rays per image. Since an image results from a 2.559~s exposure, this number translates into a fraction of 2.7\% of an image ($512\time512$ pixels) affected by cosmic rays per 1\,000~s. This number is consistent with the 1.5--3\% per 1\,000~s estimated by \citet{Riess:2002vx} for the Advanced Camera for Surveys on board \hst

\subsubsection{Identification of bad pixels}
\label{sec:badpixels}
Continuously bad pixels, in contrast with transient cosmic rays, are flagged by the WFC3 pipeline and their positions are stored in the quality information provided for each exposure. We consider that bad pixels include pixels with quality flag corresponding to bad detector pixel ($\rm flag = 4$), unstable photometric response ($\rm flag = 32$), and bad or uncertain flat value ($\rm flag = 512$).

Two bad pixels flagged as having an unstable response appear in most images right over the spectral trace at positions of $(331,279)$ and $(331,280)$. At this location, we indeed find a flux spike in most spectra. This creates an artifact in the spectrum at a wavelength of about 1.26~\micron, hence we will discard data from this spectral channel in the remaining of this analysis. Apart from this effect, a careful eye-analysis of our raw images and spectra did not reveal any significant feature that could be due to bad pixels in our extraction windows. 

\subsubsection{Spectral extraction}
\emph{Wavelength calibration.}
\label{sec:wavecalib}
The wavelength calibration is based on the determination of the spectral trace location in the grism images. This location does itself depend on the position $(x_\mathrm{ref},y_\mathrm{ref})$ of the star in the direct images. The star position is determined with a two-dimensional Gaussian fit to the direct image for each \hst\ orbit. (The variation in the centroid position of the direct image is at most 0.2~pixel in $x$ and $y$.) Once these positions are obtained, the $(x,y)$ coordinates of the spectral trace in the grism images are then given by
\begin{equation}
 y - y_\mathrm{ref} = DYDX_0 + DYDX_1 (x - x_\mathrm{ref}),
\end{equation}
where the values of $DYDX_0$ and $DYDX_1$ are taken for the $+1^\mathrm{st}$-order spectrum from \citet[][see their Table~1]{Kuntschner:2009tq}. We found that depending on the considered \hst\ orbit, the wavelength calibration can be offset by as much as $\sim$0.2 pixel. Before comparing or co-adding spectra obtained during different \hst\ orbits, it is therefore important to interpolate them on a common reference wavelength vector. We chose the first exposure of the second \hst\ orbit as our reference.

\begin{figure}[!ht]
\resizebox{\columnwidth}{!}{\includegraphics[trim=0 2cm 0 1cm]{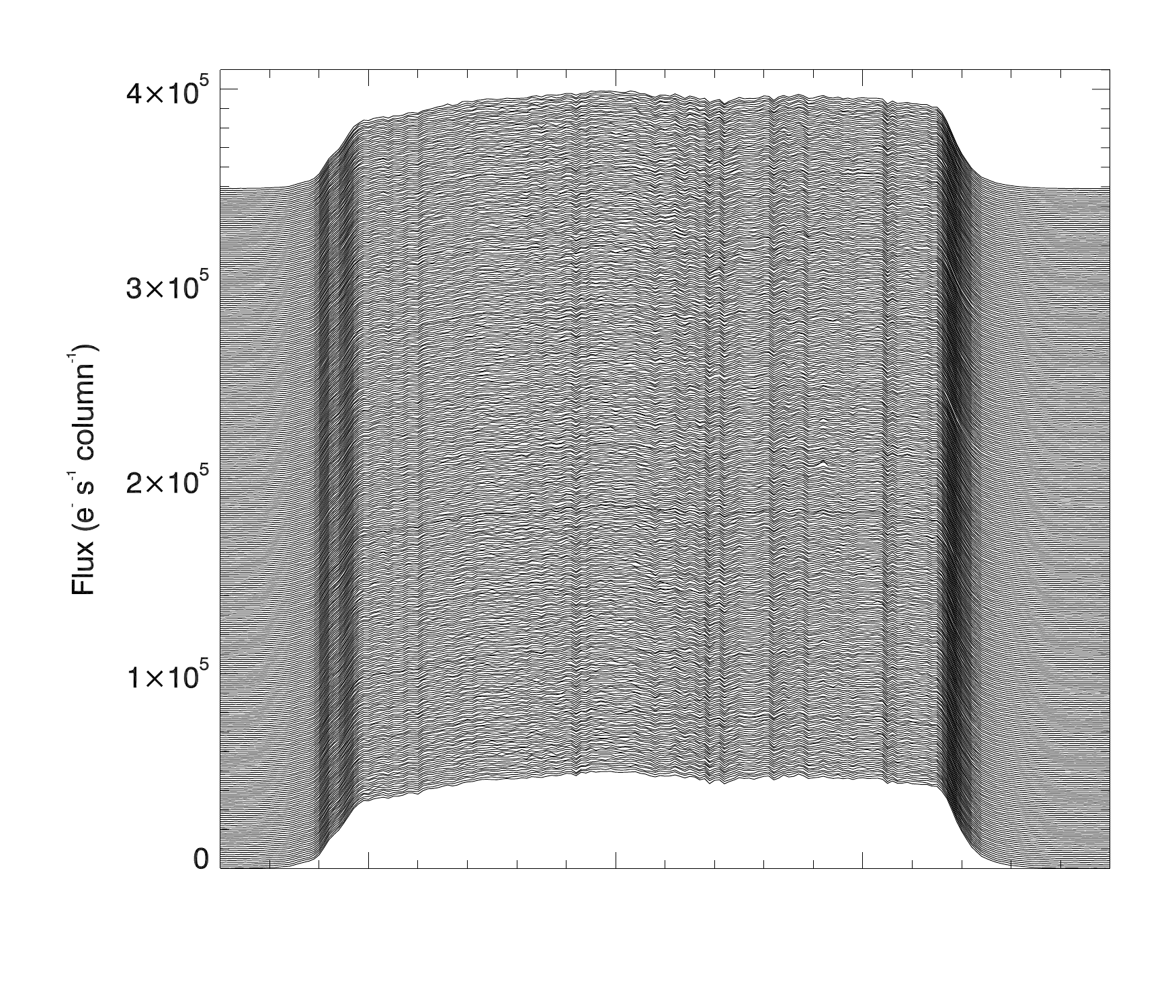}}
\resizebox{\columnwidth}{!}{\includegraphics[trim=0 0 0 1cm]{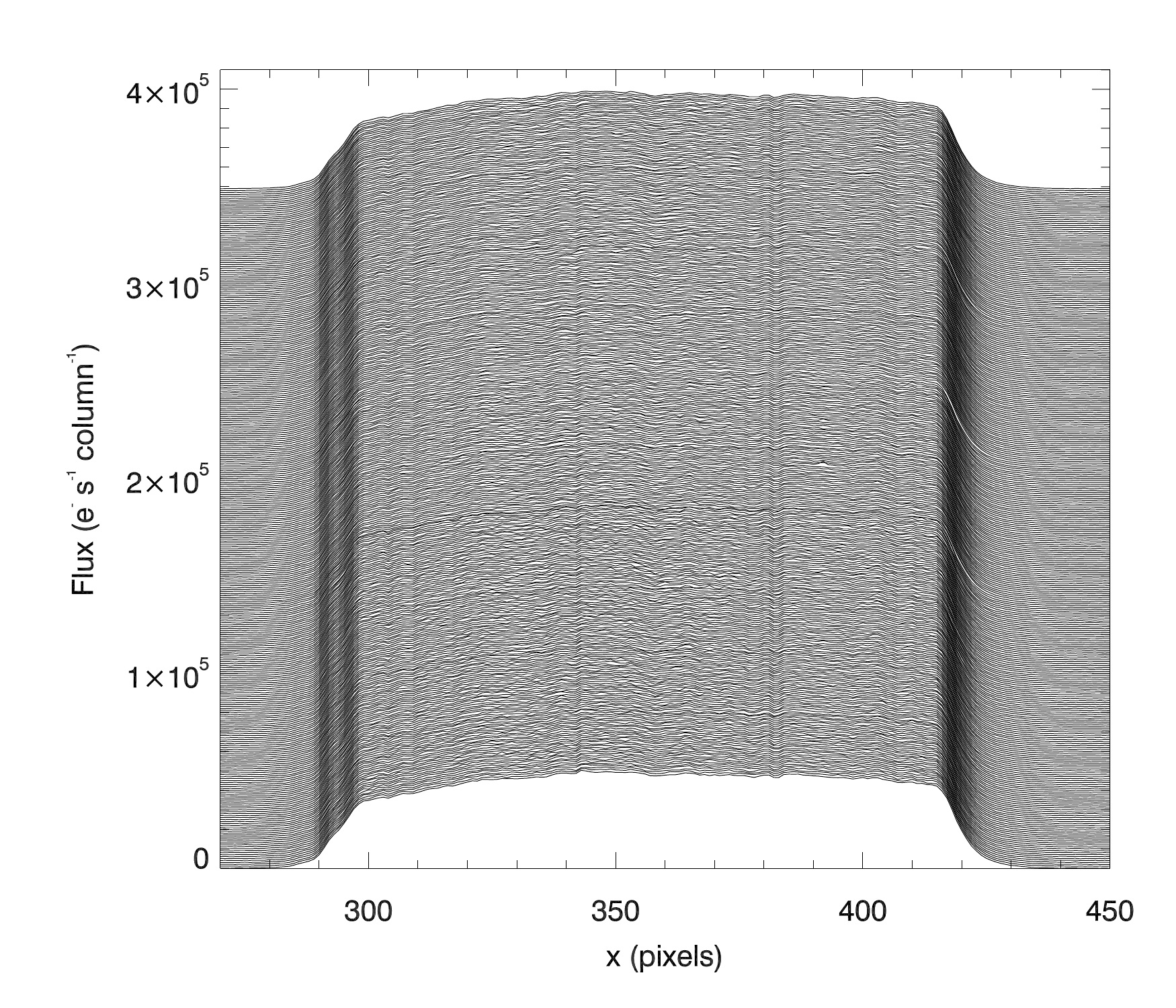}}
\caption{\label{fig:allspectra}The 350 spectra extracted from the grism images. The spectra are offset in flux for clarity. \emph{Top panel---} Spectra extracted from the raw images. \emph{Bottom panel---} Spectra extracted from the flat-fielded, background subtracted images.}
\end{figure}

\emph{Extraction of the $1^\mathrm{st}$-order spectra.}
\label{sec:rawspectra}
In each image, the raw $1^\mathrm{st}$-order spectrum is extracted along the spectral trace. For each detector column along the trace, the flux $f$ is calculated as the sum of the flux $f_i$ in all pixels located in the extraction box centred on the trace $y$ location. We found that the signal-to-noise is optimised with an extraction box width of 13 pixels. The error on the flux $\sigma_f$ is obtained as the quadratic sum of the errors ${\sigma_f}_i$ estimated for each pixel by the WFC3 pipeline. The 350 raw spectra are shown in Fig.~\ref{fig:allspectra} (top panel).

\subsubsection{Wavelength-dependent flat fielding}
\label{sec:paragraph}
\begin{figure}[!ht]
\resizebox{\columnwidth}{!}{\includegraphics{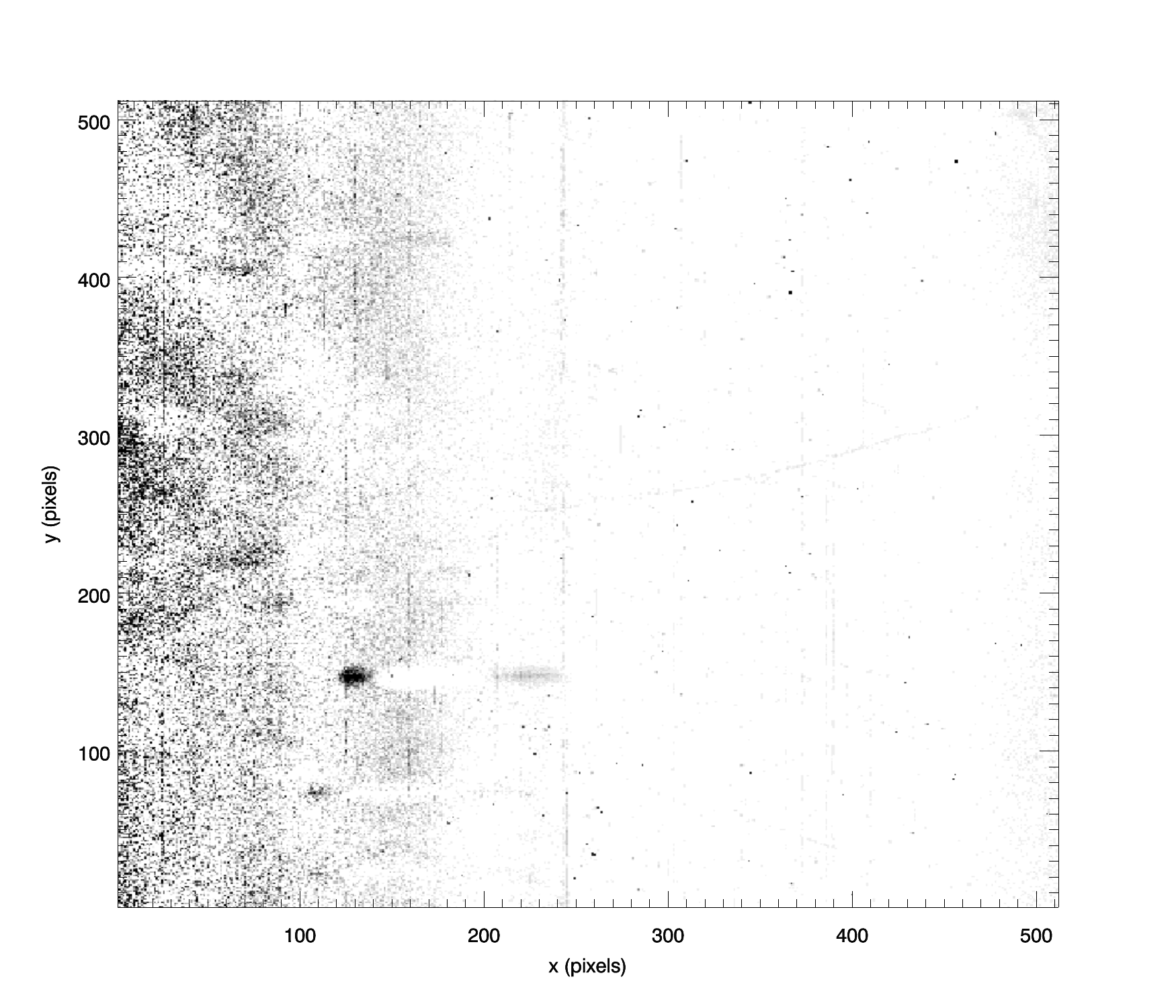}}
\caption{\label{fig:flatfield}Wavelength-dependent flat-field image. The scale is linear and ranges between $-4.2$ to $+6.4$~e$^-$~s$^{-1}$ (black to white, respectively).}
\end{figure}

The WFC3 pipeline does not apply flat-field corrections for grism images. In order to perform this task, we obtained a reference flat-field cube for WFC3/G141 observations.\footnote{\label{fn:wfc3url}\url{www.stsci.edu/hst/wfc3/analysis/grism_obs/calibrations/wfc3_g141.html}} 
The coefficients from the cube are used to calculate the flat-field image following \citep[][Sect.~6.2]{Kummel:2011up}. The flat-field image contains a handful of zeros, that we replace by the median of the 8 neighbouring pixels to avoid troubles at the division level. The flat-field image is shown in Fig.~\ref{fig:flatfield}. All grism images are then divided by the wavelength-dependent flat-field images.

\subsubsection{Background subtraction}
For each flat-fielded exposure, the background is extracted one hundred pixels above and below the $1^\mathrm{st}$-order spectrum, along traces parallel to the spectral trace. The width of the extraction window (along the $y$ axis) at both locations is equal to the width of the spectral extraction window. The background value along the spectral trace is taken as the average of the background values extracted above and below the spectrum, before being subtracted to it. The background-subtracted, flat-fielded spectra are shown in Fig.~\ref{fig:allspectra} (bottom panel).

\subsubsection{Flux calibration}
\label{sec:fluxcal}
\begin{figure}[!ht]
\resizebox{\columnwidth}{!}{\includegraphics{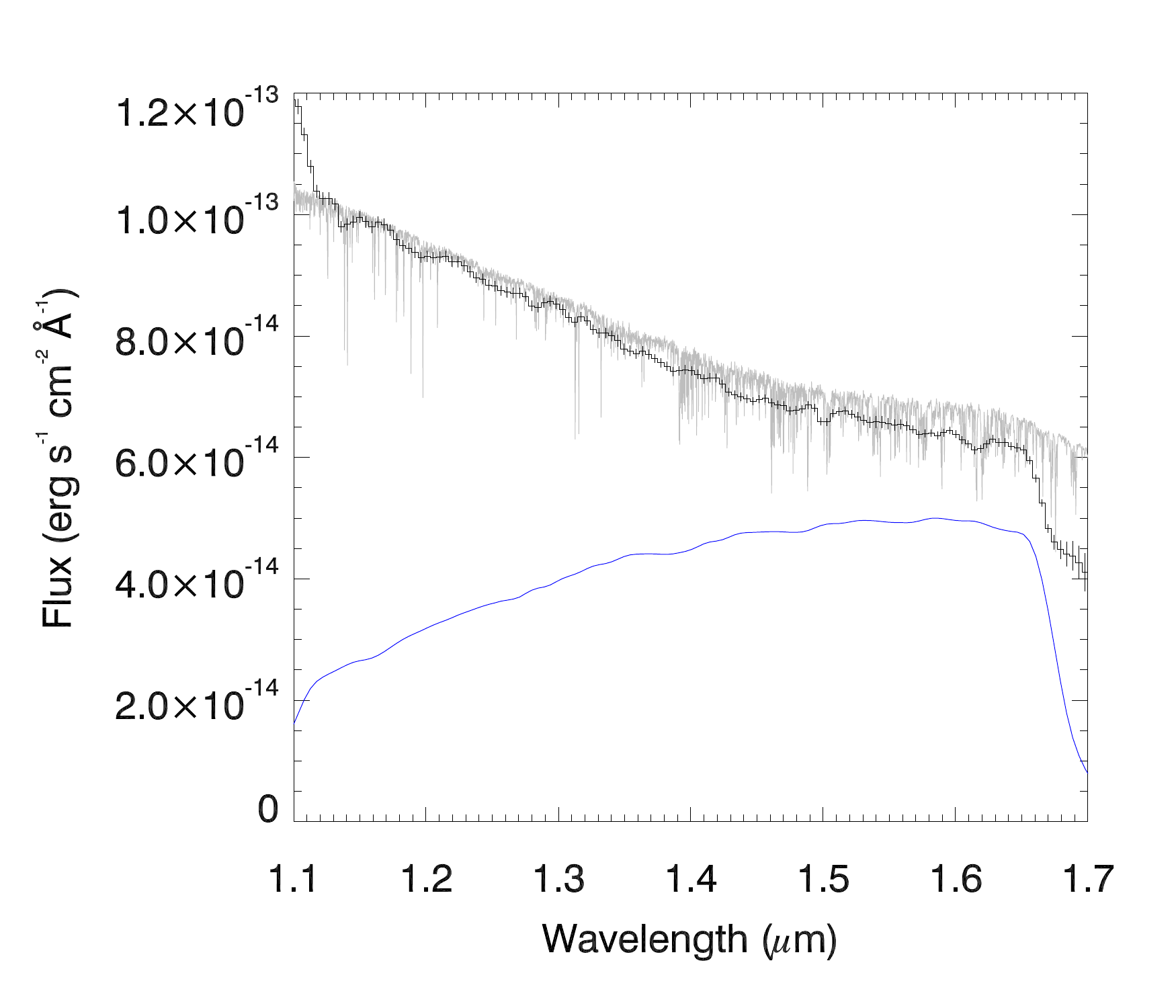}}
\caption{\label{fig:calspectrum}Mean flux-calibrated spectrum of GJ~3470 (black line). A PHOENIX model spectrum with $T_\mathrm{eff}=3\,600$~K and $\log g=4.5$ is shown in grey. Finally, the WFC3/G141 sensitivity is shown by the blue curve (arbitrary units).}
\end{figure}

The measurements reported here are essentially differential, therefore a flux calibration is not needed to calculate the transmission spectrum of the planet. However, it is useful to perform the flux calibration to compare our spectrum of GJ~3470 to model spectra and determine important stellar properties, such as the limb-darkening profile of the stellar surface.
A mean uncalibrated GJ~3470 spectrum is calculated as the average of the 350 background-subtracted, flat-fielded spectra. The flux-calibrated, mean spectrum is obtained by dividing the mean uncalibrated spectrum by the wavelength-dependent G141 sensitivity, which is obtained from the WFC3/G141 calibration and reference files web page\footnote{See footnote~\ref{fn:wfc3url}.} and interpolated to our wavelength solution. The mean flux-calibrated spectrum is shown in Fig.~\ref{fig:calspectrum}, together with a stellar atmosphere PHOENIX model from the G\"ottingen Spectrum Library\footnote{\url{http://phoenix.astro.physik.uni-goettingen.de/?page_id=82}} \citep{Husser:2013ca}. The model parameters are a stellar effective temperature $T_\mathrm{eff}=3\,600$~K, a surface gravity $\log g=4.5$, and a metallicity $\rm [Fe/H]=0.0$. These values are chosen as close as the grid of PHOENIX models allows to the stellar parameters derived by \citet{Demory:2013hv}: $T_\mathrm{eff} = 3\,600\pm100$~K, $\log g = 4.66\pm0.03$, and $\rm [Fe/H] = 0.2\pm0.1$. By rescaling the model spectra to a distance of 33~pc -- somewhat farther than the values reported by \citet[][25.2~pc]{Bonfils:2012gb} and \citet[][30.7~pc]{Demory:2013hv} -- we obtain a reasonable match to the measured spectrum.

\subsubsection{Raw white light curve}
The white light photometric time series is obtained by integrating the background-subtracted, flat-fielded stellar spectrum between 1.15 and 1.65~\micron, for each exposure.

\begin{figure}[!h]
\resizebox{\columnwidth}{!}{\includegraphics[trim=0 1.7cm 0 0, clip=true]{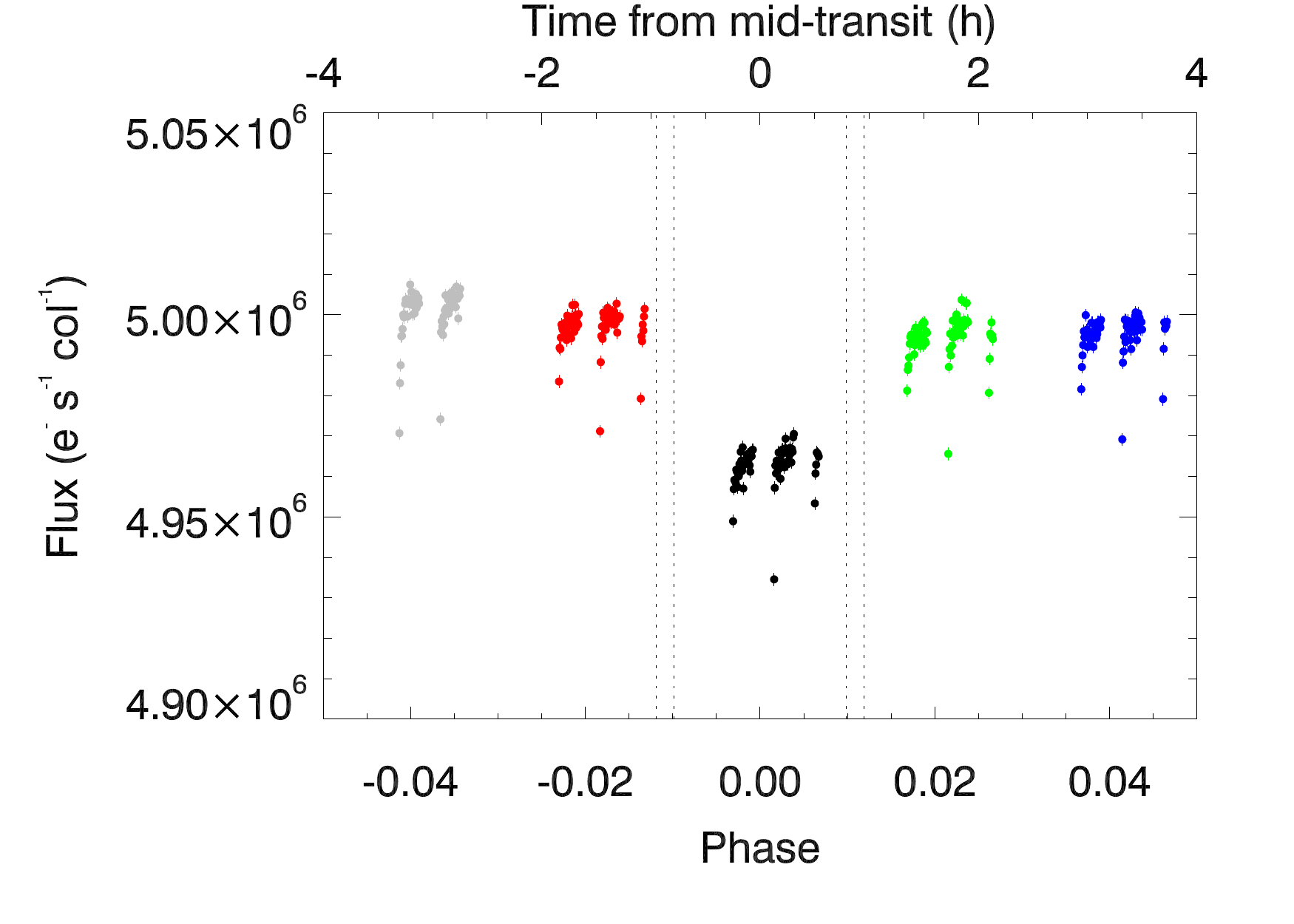}}
\resizebox{\columnwidth}{!}{\includegraphics[trim=0 0 0 0.8cm, clip=true]{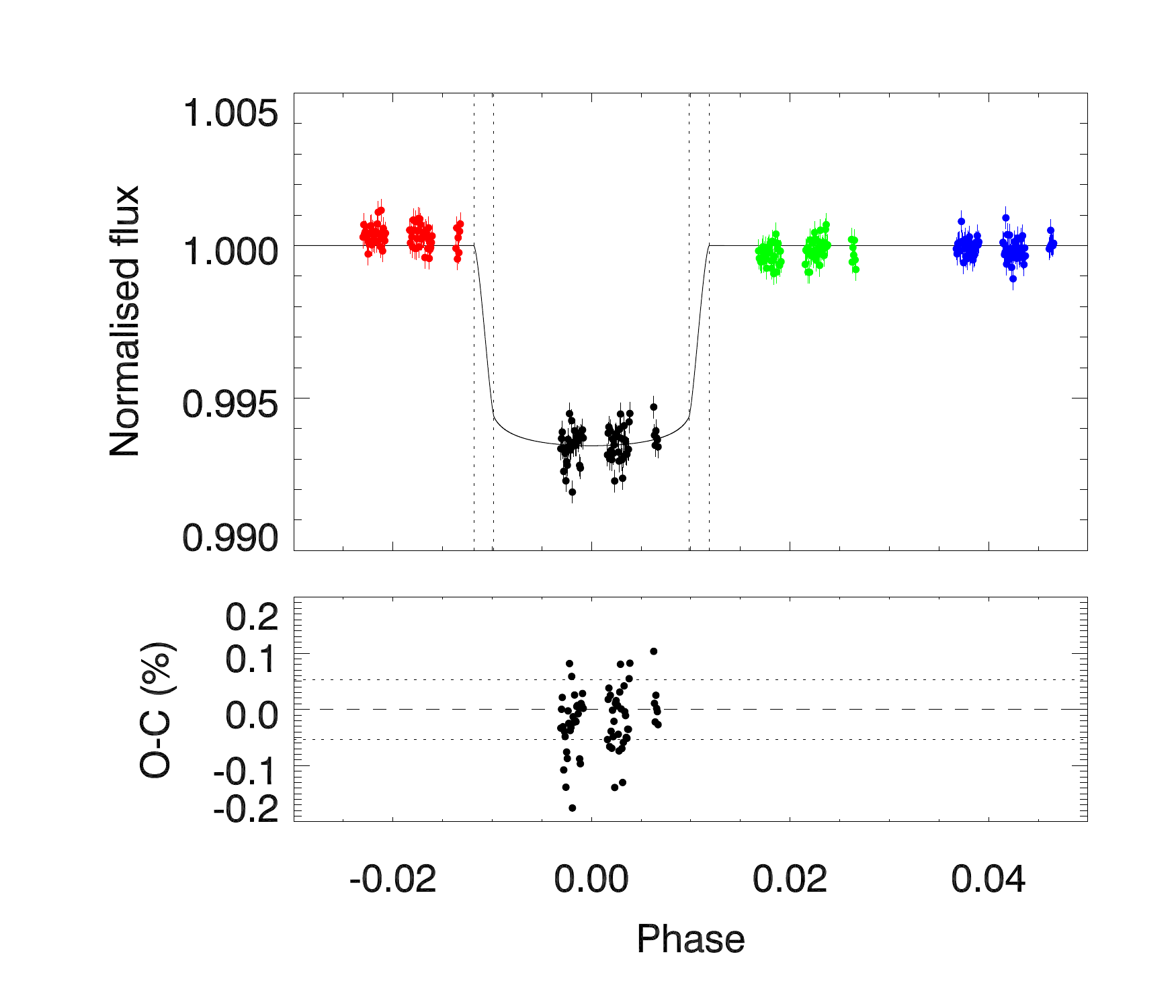}}
\caption{\label{fig:whitelc}\emph{Top panel---} Raw white light curve of GJ~3470. Photometric points obtained during each \hst\ orbit are shown in a different colour. The transit contacts of GJ~3470b are indicated by vertical dashed lines. \label{fig:ootwhitelc} \emph{Middle panel---} Corrected white light curve of GJ~3470. The systematics in the in-transit (third) \hst\ orbit (black dots) have been corrected by dividing the fluxes by a correction template, built as the mean of the second (red dots) and fourth (green dots) \hst\ orbits (which are shown here also divided by this correction template). The black curve is the best fit to the corrected white light curve. \emph{Bottom panel---} Observed minus calculated values ($\rm O-C$). The two horizontal dotted lines indicate the standard deviation of the residuals in the in-transit (third) \hst\ orbit.}
\end{figure}

\begin{figure}[!ht]
\resizebox{\columnwidth}{!}{\includegraphics{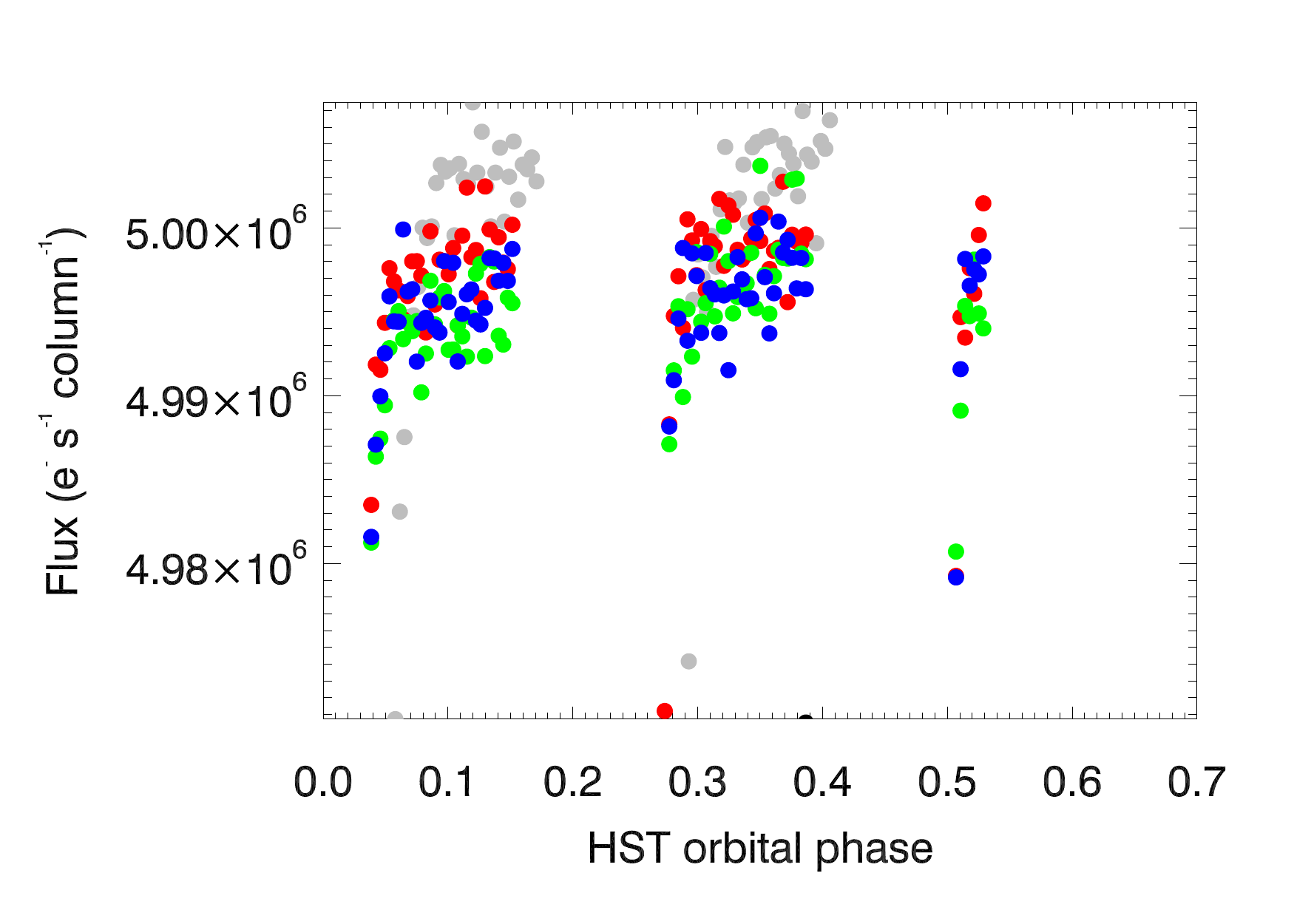}}
\caption{\label{fig:orbitlc}Raw light curve of GJ~3470 phase-folded over one \hst\ orbit (96~min). The different colours refer to the different orbits, following the same convention as in Fig.~\ref{fig:whitelc}.}
\end{figure}

\begin{figure}[!ht]
\resizebox{\columnwidth}{!}{\includegraphics{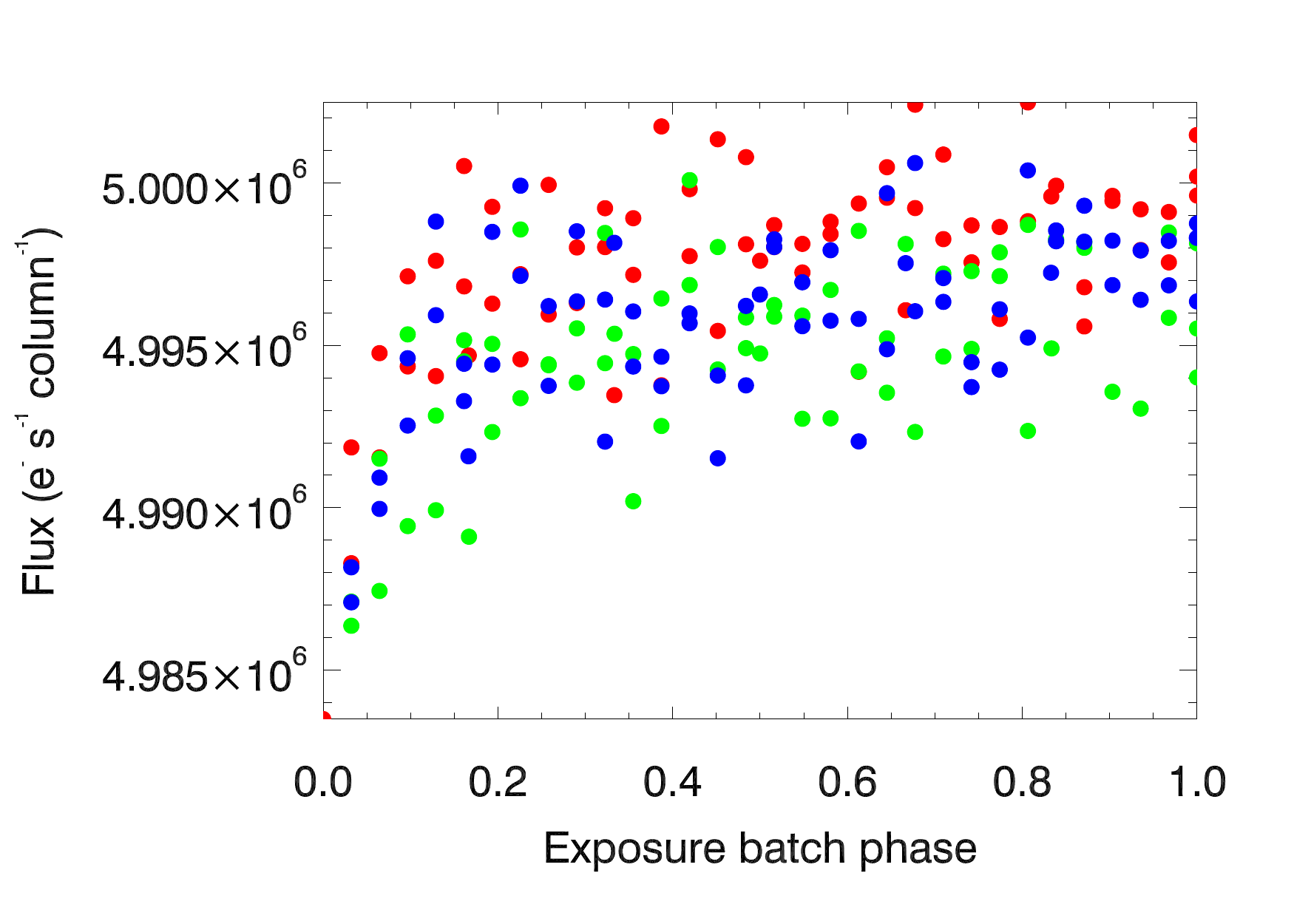}}
\caption{\label{fig:batchlc}Raw light curve of GJ~3470 phase-folded over a batch of exposures. The different colours refer to the different orbits, following the same convention as in Fig.~\ref{fig:whitelc}. The first orbit is not shown here.}
\end{figure}

 The raw white light curve is shown in Fig.~\ref{fig:whitelc}. The raw light curve is affected by systematic effects, among which the most salient is a rapid flux rise at the beginning of each batch of exposures. This pattern, known as the fish hook \citep{Deming:2013ge}, is reproducible between one exposure batch to the other, for all \hst\ orbits, as shown by Figs.~\ref{fig:orbitlc} and~\ref{fig:batchlc}, where the white light curve is phase-folded over one \hst\ orbit and one batch of exposures, respectively.

\subsubsection{Raw chromatic light curves}
In order to measure the planetary radius at different wavelengths, hence producing a transmission spectrum, it is necessary to obtain photometric time series at different wavelengths. The classical approach for WFC3/G141 data is first to bin the spectrum, then to obtain chromatic light curves for each wavelength bin. In fact, efficiently extracting the full set of transit parameters (planet-to-star radius ratio, system scale, impact parameter, limb-darkening coefficients) for each wavelength bin demands high-signal-to-noise light curves. Higher signal-to-noise ratios require larger bins, which can inevitably lead to ignore any wavelength-dependent systematic effect. As we will see, however, our analysis needs only to extract the wavelength-dependent planet-to-star radius ratio ($R_p/R_\star$). For this purpose, it is sufficient (in terms of signal-to-noise ratio) to obtain a light curve for each (unbinned) spectral channel and fit a transit model to each light curve to obtain the $R_p/R_\star$ value at this wavelength. The transmission spectrum ($R_p/R_\star$ vs.\ $\lambda$) can then be rebinned to increase the signal-to-noise. In our case, we found that both approaches lead to similar results, although the uncertainties in the final transmission spectrum appear to be closer to a Gaussian distribution when data binning occurs in the last steps of the reduction. This approach is the one we will present in the following.

\subsection{Data analysis}

\subsubsection{Stellar limb-darkening}
\label{sec:ld}
An important issue with the data set discussed here is that the time series provide no information about the transit egresses. Therefore, the limb-darkening profile of the star cannot be significantly constrained from fits to the white or chromatic light curves. Meanwhile, we expect not to be overly sensitive to any remaining uncertainties in the limb-darkening correction, since the transit is not grazing and the stellar spectrum at those wavelengths is relatively featureless compared to that of a later-type star such as GJ~1214 \citep[where a strong water absorption occurs][]{Berta:2012ff}. We were nevertheless careful in relying onto an adapted limb-darkening law and a realistic set of priors for the limb-darkening coefficients. For that purpose, we modelled the stellar atmosphere with the PHOENIX model \citep{Husser:2013ca} used in Sect.~\ref{sec:fluxcal} for the flux calibration, with stellar parameters $T_\mathrm{eff} = 3\,600$~K, $\log g = 4.5$, and $\rm [Fe/H] = 0.0$. 

\begin{figure}[!ht]
\resizebox{\columnwidth}{!}{\includegraphics{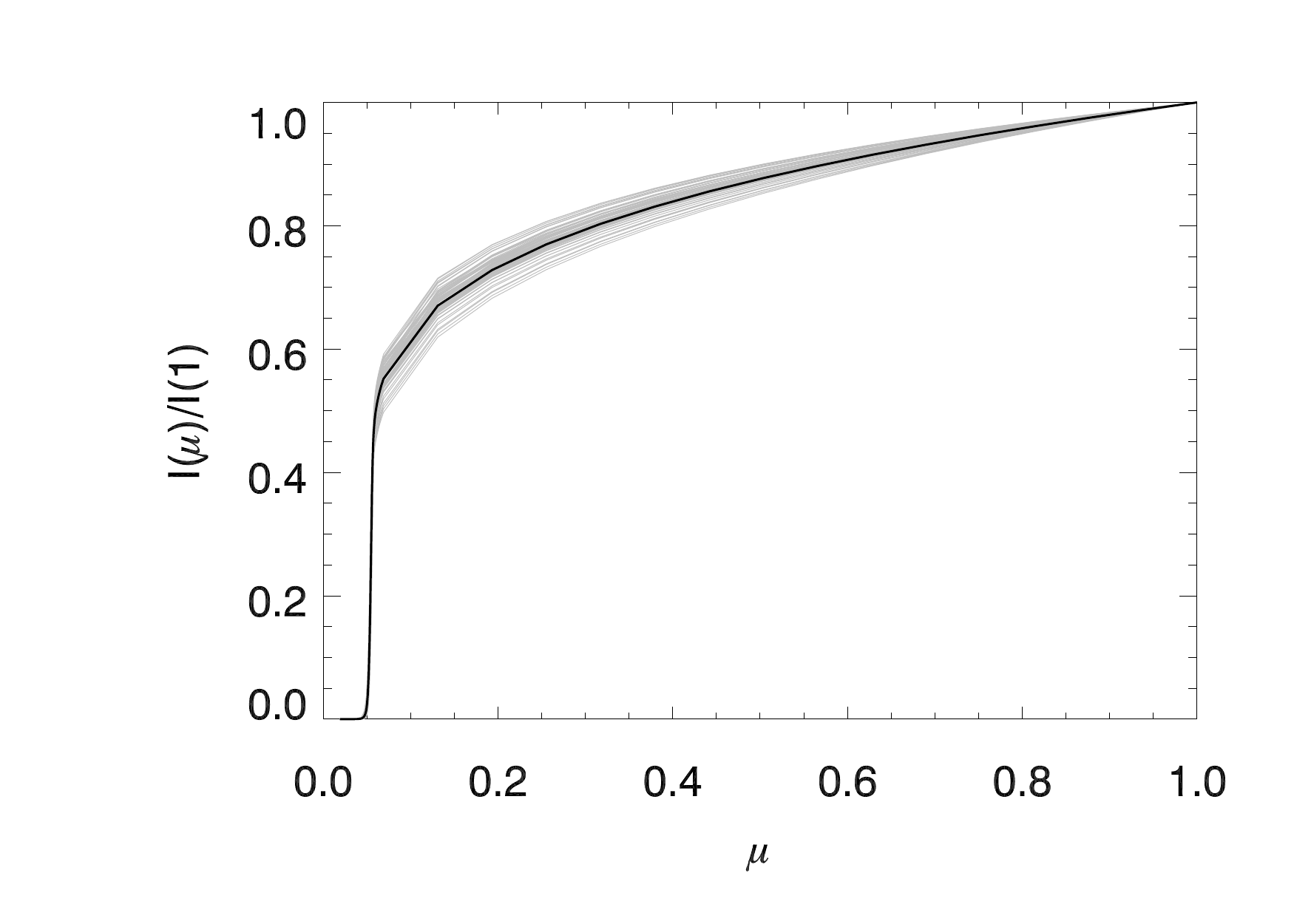}}\caption{\label{fig:ldini}Limb-darkening profiles for the PHOENIX stellar atmosphere model ($T_\mathrm{eff}=3\,600$~ K, $\log g = 4.5$, $\rm [Fe/H] = 0.0$) The grey curves are the profiles obtained for 78 wavelength values across the WFC3/G141 band pass. The black line is the limb-darkening profile integrated over this band pass.}
\end{figure}

\begin{figure}[!ht]
\resizebox{\columnwidth}{!}{\includegraphics{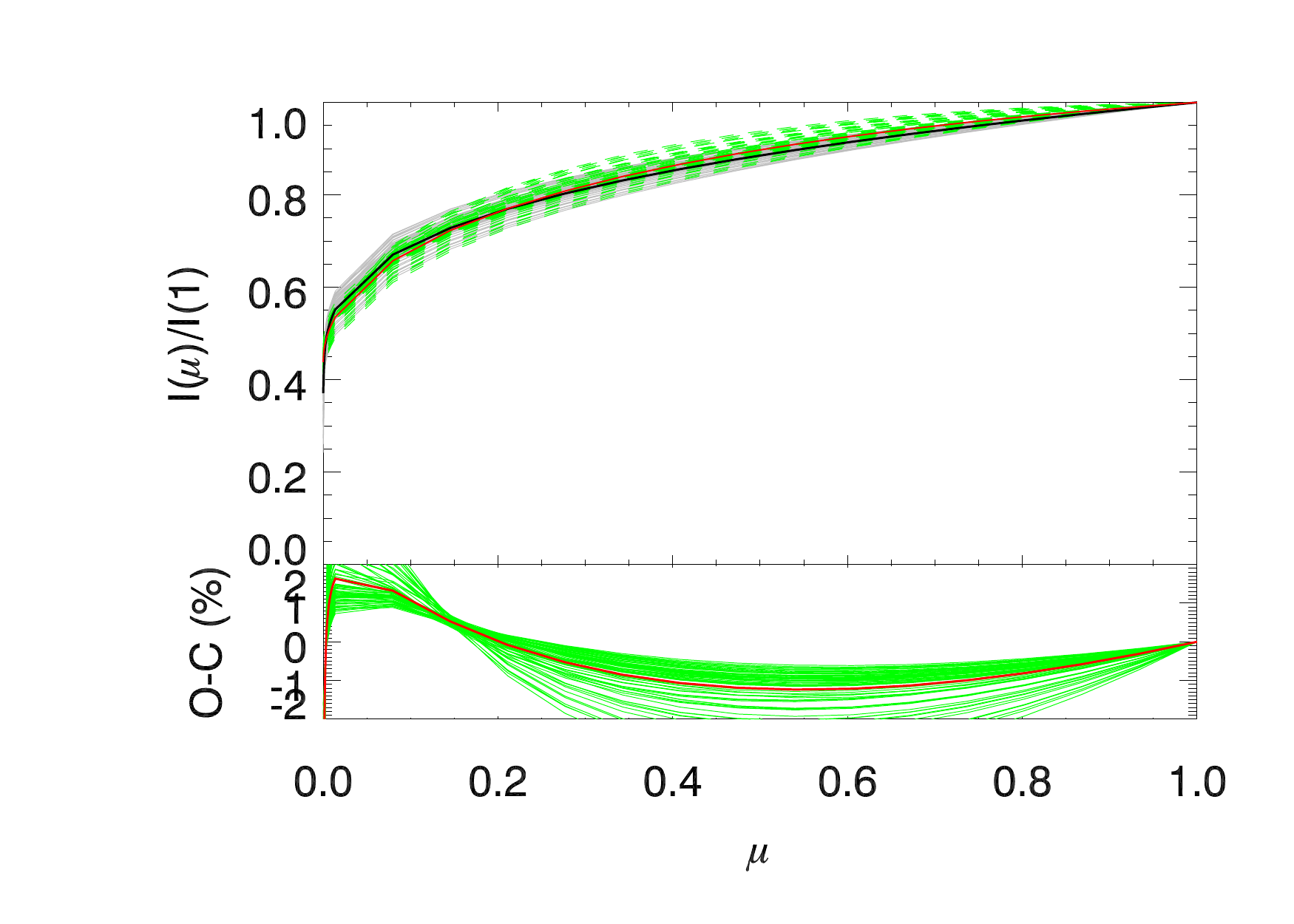}}\caption{\label{fig:ldcropped}\emph{Top panel---} Same as in Fig.~\ref{fig:ldini}, except that the stellar surface ($\mu=0$) is taken where the specific intensity has dropped by a factor of $\exp{(1)}$. The green dashed curves are the best fits to the wavelength-dependent profiles in grey. The solid red line is the best fit to the integrated limb-darkening profile (black line). \emph{Bottom panel---} Observed minus calculated values of the wavelength-dependent profiles (green curves) and the integrated profile (red curve).}
\end{figure}

The specific intensity profile table was retrieved from the G\"ottingen Spectrum Library. It contains the stellar intensity values for 78 angles ($\mu$) and for 25\,500 wavelengths. It thus allows the retrieval of 25\,500 limb-darkening profiles, one for each wavelength bin. We extracted the profiles corresponding to the wavelength range covered with WFC3/G141 and binned them in wavelength by a factor of 100, yielding 77 limb-darkening profiles with 78 angle values each. These profiles are shown in Fig.~\ref{fig:ldini} with the profile integrated over the whole WFC3/G141 band pass. These profiles show a steep decrease below $\mu \sim 0.1$, which cannot be easily modelled with the classical non-linear, quadratic, or square-root limb-darkening laws. This is because the modelled stellar radius ($\mu = 0$) is chosen to encompass all the stellar flux at all wavelengths provided by the PHOENIX model. For most wavelengths in the visible and near-infrared, the stellar surface will be found at higher values of $\mu$ (T.-O.~Husser, private communication). As also done by \citet{Berta:2012ff}, we thus cropped the profiles in Fig.~\ref{fig:ldini} and considered that the stellar surface ($\mu = 0$) is reached at the angle where the central intensity $I(1)$ has decreased by a factor of $\exp{(1)}$. This leaves 37 values of $\mu$ per profile. The resulting 77 profiles are plotted in Fig.~\ref{fig:ldcropped} (top panel). 

The profiles were fit with a square-root law, where the last two coefficients $c_3$ and $c_4$ of the non-linear limb-darkening law are set to zero, yielding a profile described by 
\begin{equation}
\frac{I(\mu)}{I(1)} = 1 - c_1(1-\mu^\frac{1}{2}) - c_2(1-\mu).
\end{equation}
The residuals are shown in the bottom panel of Fig.~\ref{fig:ldcropped}. A quadratic limb-darkening law (where $c_1=c_3=0$) would provide a marginally less-good fit, although this would not affect our results. We checked that the results do not depend on the chosen stellar atmosphere model, by trying out models with different effective temperatures in the $\pm100$~K uncertainty range derived by \citep{Demory:2013hv}, and by varying the surface gravity by $\pm0.5$.

\begin{figure}[!ht]
\resizebox{\columnwidth}{!}{\includegraphics{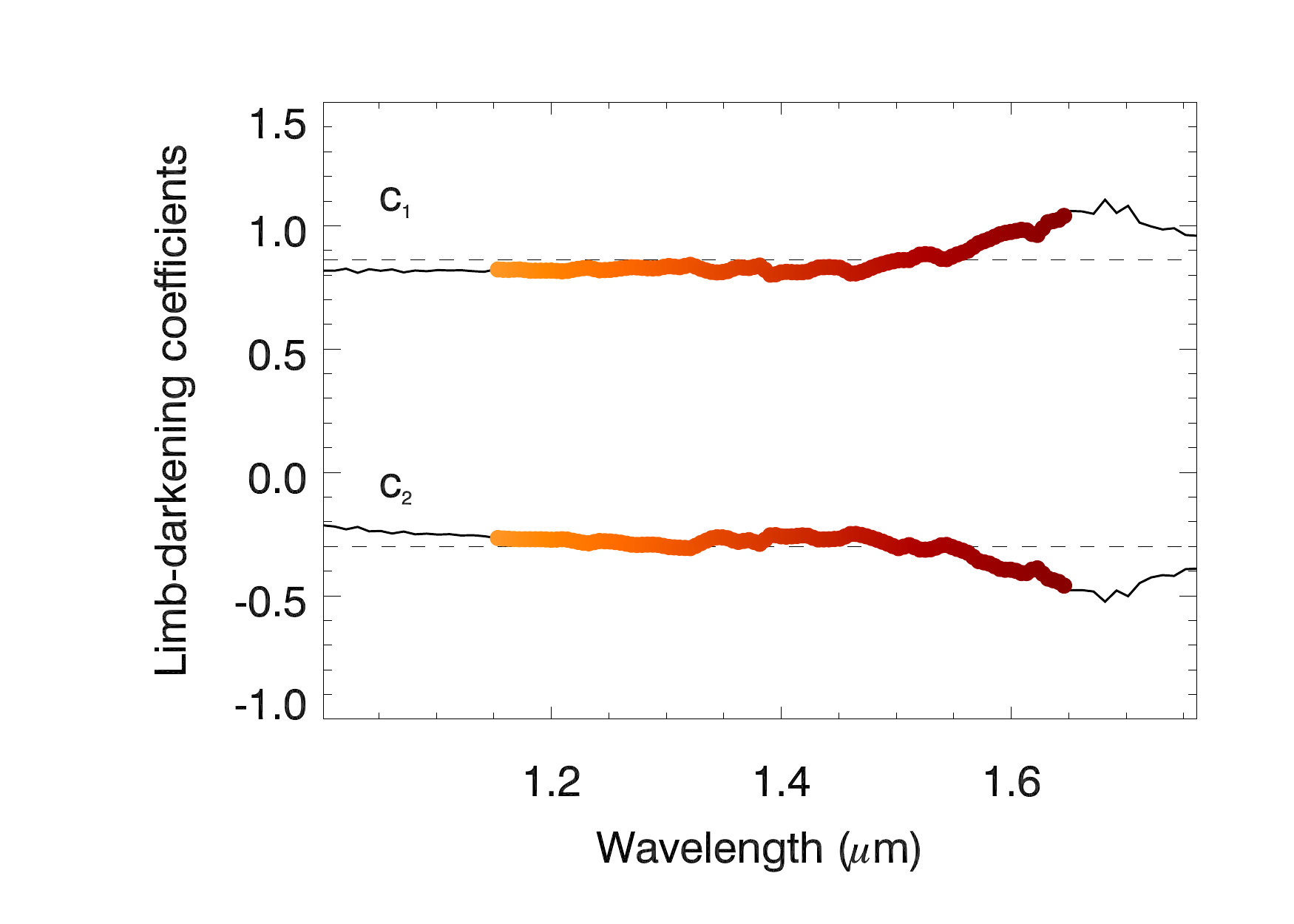}}\caption{\label{fig:ldcoefs}Wavelength-dependent limb-darkening coefficients $c_1$ (top curve) and $c_2$ (bottom curve) for the square-root law. The horizontal dashes show the values of $c_1$ and $c_2$ obtained from the fit to the limb-darkening profile integrated over the whole WFC3/G141 band pass. The coloured dots are the values of the coefficients interpolated over each wavelength bin.}
\end{figure}

The coefficients of the square-root limb-darkening law ($c_1$ and $c_2$) are represented as a function of wavelength in Fig.~\ref{fig:ldcoefs}. The values of the coefficients for each wavelength channel are then obtained through interpolation.

\subsubsection{Correction of instrumental systematics with the out-of-transit data}
\label{sec:oot-divide}
We experimented with several methods to correct for the systematic effects in the white and chromatic light curves. The first method is based on the reasonable assumption that the systematics are reproducible between successive \hst\ orbits within one \hst\ visit. It consists in dividing the in-transit measurements by a \hst-orbital-phase-dependent correction template calculated as the mean of the out-of-transit (OOT) measurements obtained before and after transit. This is the so-called \texttt{divide-oot} procedure \citep{Berta:2012ff,Wilkins:2014ga,Ranjan:2014hv}. The first \hst\ orbit in a visit is traditionally dismissed at this stage because of enhanced systematics resulting from telescope settling and readjusting to a new pointing \citep[e.g.,][]{Wilkins:2014ga}. In our case, first-orbit exposures are also slightly offset in phase from the other \hst\ orbits, as can be seen in Fig.~\ref{fig:orbitlc}, because of the guide star acquisition procedure in the first orbit. Furthermore, no third exposure batch was recorded during the first orbit because of this acquisition. We thus discarded the first \hst\ orbit for the remaining of this analysis. Visit-long effect, such as an increasing ramp of the exposure levels, could be problematic when averaging structures on a per-\hst-orbit basis \citep{Wilkins:2014ga}. In our case, however, no significant ramp is observed in the raw white light curve (Fig.~\ref{fig:whitelc}, top panel), making it possible to construct the correction template as the average of \hst\ orbits~2, 4, and~5.
 
The errors on the correction template is calculated as a function of time, as the quadratic mean of the errors on averaged \hst\ orbits. They are propagated to the corrected (divided) flux of the in-transit measurements. The results can be seen in Fig.~\ref{fig:ootwhitelc} for the white light curve. In this figure, the corrected data obtained before and after transit are shown; however they are not used in the fitting procedure described below. In fact, the correction template is built from these out-of-transit data, therefore these corrected \hst\ orbits do not contain any information regarding any remaining systematic effects, contrary to the corrected in-transit (third) \hst\ orbit.

\emph{White light curve.} We adjusted the corrected white light curve with a transit model based on the \texttt{occultnl} set of routines by \citet{Mandel:2002bb} and a least-square minimisation algorithm based on the Levenberg-Marquardt technique (C.~Markwardt's \texttt{MPFIT}\footnote{\url{http://cow.physics.wisc.edu/~craigm/idl/idl.html}}). The only free parameter is the planet-to-star radius ratio ($R_p/R_\star$). The impact parameter ($b$) and the system scale ($a_p/R_\star$) have been fixed to the values determined by \citet{Demory:2013hv} from the \emph{Spitzer} 4.5~\micron\ light curve ($b=0.4$, $a_p/R_\star=13.46$). The limb-darkening coefficients of the square-root law are fixed to the values obtained for the specific intensity profile integrated over the whole WFC3/G141 band pass (see Sect.~\ref{sec:ld}), $c_1 = 0.862$ and $c_2 = -0.300$. 

\emph{Chromatic light curves.} The same procedure used for the white light curve was carried out on the raw chromatic light curves: a correction template was calculated for each light curve in order to deal with any potential wavelength-dependent effects. We found, however, that calculating a correction template for each spectral channel or using the correction template calculated from the white light curve throughout all channels makes little difference in the final result. Since the white light curve correction template has a higher signal-to-noise ratio than the per-channel templates, choosing the former propagates less noise into the transmission spectrum. The corrected chromatic light curves are shown in Fig.~\ref{fig:ootchromlc}. Each light curve was fitted with the transit model described above. Here again, we only set the planet-to-star radius ratio as a free parameter (since we are only interested in the wavelength-dependent variation of this parameter). The impact parameter and system scale are fixed to the same values as for the white light curve fit. The limb-darkening coefficients are fixed to the values determined in Sect.~\ref{sec:ld} and plotted in Fig.~\ref{fig:ldcoefs}. We verified that the error bars on the radius ratios yielded by the fitting procedure are similar to those obtained through a $\Delta \chi^2$ analysis. For each spectral channel, we rescaled the uncertainty obtained by the value of the reduced $\chi^2_\nu$.

\subsubsection{Correction of instrumental systematics with the white light curve}
\label{sec:differential-correction}
\citet{Wilkins:2014ga} propose a differential method to obtain a transmission spectrum from WFC3/G141 stare mode data. The method consists in, just after spectral extraction, dividing the flux in each spectral channel at a given time by the white flux integrated over the whole spectrum at the same time (albeit not including the considered spectral channel). This has the advantage to correct for any time correlation in the data, not only including the hook effect (which mostly depends on \hst\ orbital phase for a given flux level per pixel), but also other possible effects such as the motions of the spectrum over time caused by the spacecraft jitter. The disadvantage of the method is to produce normalised light curves which are not transit light curves any more (the white transit light curve has been corrected out), and thus are not as straightforward to model and adjust as in the \texttt{divide-oot} method. However, the differential spectral signature of the planetary atmosphere $\delta F$ is conserved in the ratio between the in-transit \hst\ orbit and the out-of-transit \hst\ orbits. Since we do not observe the transit egresses, it is tempting simply to integrate all in-transit exposures and divide by the integration of all out-of-transit exposures, $\delta F(\lambda) = F_\mathrm{in}/F_\mathrm{out}$. This is done at the cost of assuming that limb-darkening has no strong effect on our results (which is the case here). The differential measurements obtained with this method are added to the white light curve radius ratio to yield the chromatic radius ratio measurements, $k(\lambda) = \sqrt{1 - \delta F(\lambda) + (R_p/R_\star)^2}$. The uncertainties $\sigma_{k(\lambda)}$ are propagated following 
\begin{equation}
\sigma_{k(\lambda)} = \frac{\sqrt{\delta F(\lambda)^2 + \left(2 R_p/R_\star \sigma_{R_p/R_\star}\right)^2}}{2 k(\lambda)}.
\end{equation}

We found that this method yields similar results to the \texttt{divide-oot} method (Sect.~\ref{sec:oot-divide}). It does so, remarkably, without performing any light curve fitting or limb-darkening correction (see Fig.~\ref{fig:transpectrum-oot}).

\section{Results}
\label{sec:results}

\subsection{Broad-band near-infrared radius of GJ~3470b}
\begin{figure}[!ht]
\resizebox{\columnwidth}{!}{\includegraphics{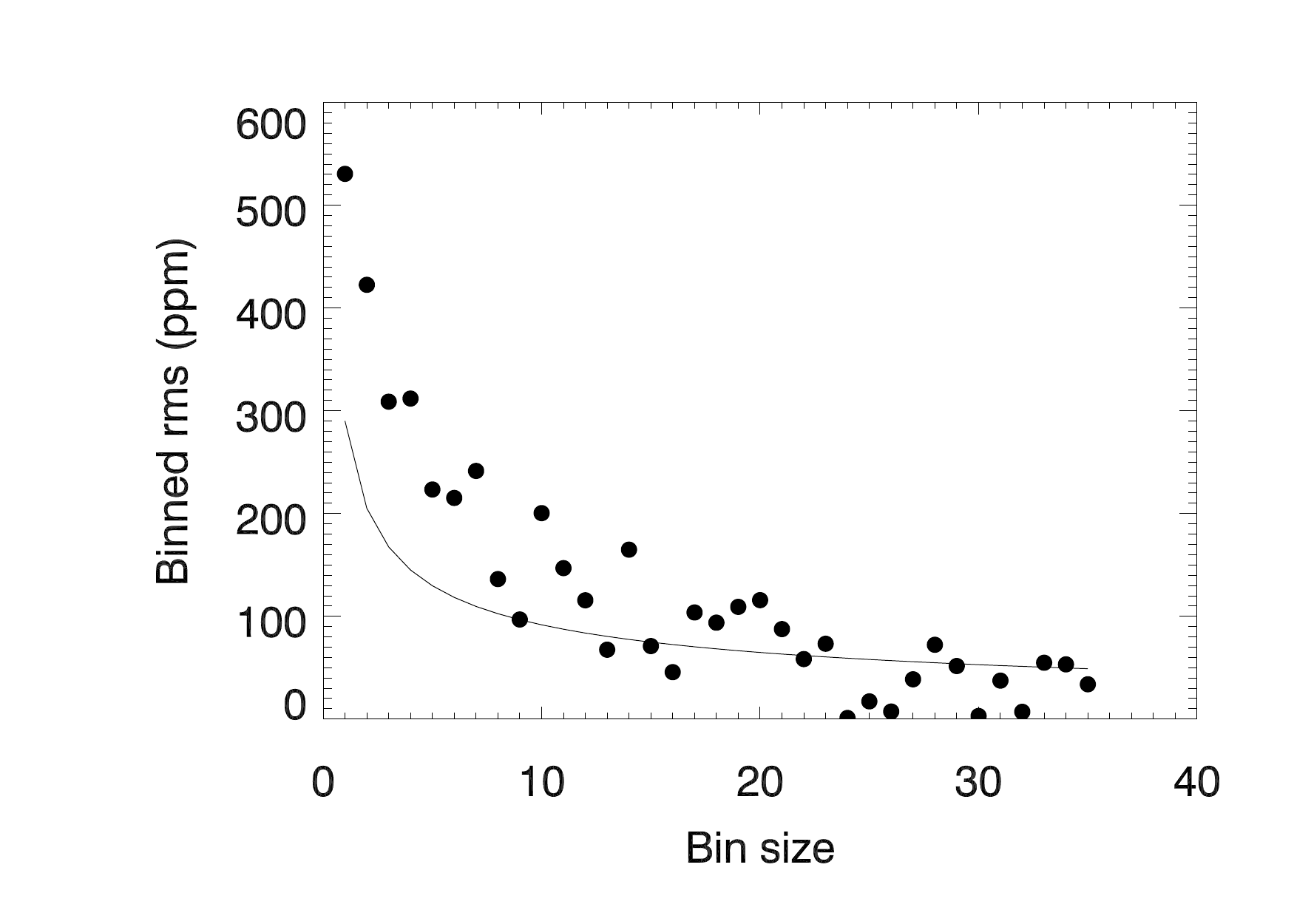}}
\caption{\label{fig:oot-white-residual-bins}Binned rms of the residuals from the white light curve in Fig.~\ref{fig:whitelc} obtained with the \texttt{divide-oot} method. The plain curve show how the binned rms should behave if they were Gaussian ($\propto 1/\sqrt{N}$). The theoretical shot noise of the unbinned residuals is $\sim300$~ppm.}
\end{figure}

The best fit to the white light curve obtained from the \texttt{divide-oot} method (see Fig.~\ref{fig:ootwhitelc}, middle panel) is obtained for a planet-to-star radius ratio of $R_p/R_\star = (7.848\pm0.037)$\%. The quoted uncertainty has been rescaled to take into account the $\chi^2$ value of 140 for 70 degrees of freedom (reduced $\chi^2_\nu=2.0$). The root mean square (rms) of the residuals, which are plotted in the bottom panel of Fig.~\ref{fig:whitelc}, is 526~ppm. Figure~\ref{fig:oot-white-residual-bins} shows that binning the residual rms by an increasing factor $N$ yields values compatible with what would be obtained assuming Gaussian noise dominated by shot noise ($\propto 1/\sqrt N$). At low $N$, however, the observed noise is above the shot noise (by a factor of 1.75): this is possibly caused by remaining systematics or correlated noise between adjacent spectral channels. 

The broad-band radius ratio corresponding to the WFC3/G141 band pass and derived from our measurements is in agreement ($1.4\sigma$) with, and of similar precision to the photometric value of $(7.798^{+0.046}_{-0.045})\%$ obtained at 4.5~\micron\ with \emph{Spitzer} \citep{Demory:2013hv}. Adopting the stellar radius of $R_\star = 0.568\pm0.037$ derived by these authors, we find that GJ~3470b has a near-infrared radius of $R_p = 4.86\pm0.32$~\Rearth. 

The aim of this work is to provide constraints on the atmosphere of the planet. To this purpose, it is already interesting to compare the broad-band measurements obtained with \hst\ at 1.1--1.7~\micron\ and \emph{Spitzer} at 4.5~\micron : the fact that the measurements are compatible to $1.4\sigma$ hint at a flat transmission spectrum between these wavelengths. In the section below, we explore the spectroscopic measurements to determine whether this is indeed the case.

\subsection{Transmission spectrum of GJ~3470b from 1.1 to 1.7~\micron}

\begin{figure}[!ht]
\resizebox{\columnwidth}{!}{\includegraphics[trim=2cm 2.2cm 0 0, clip=true]{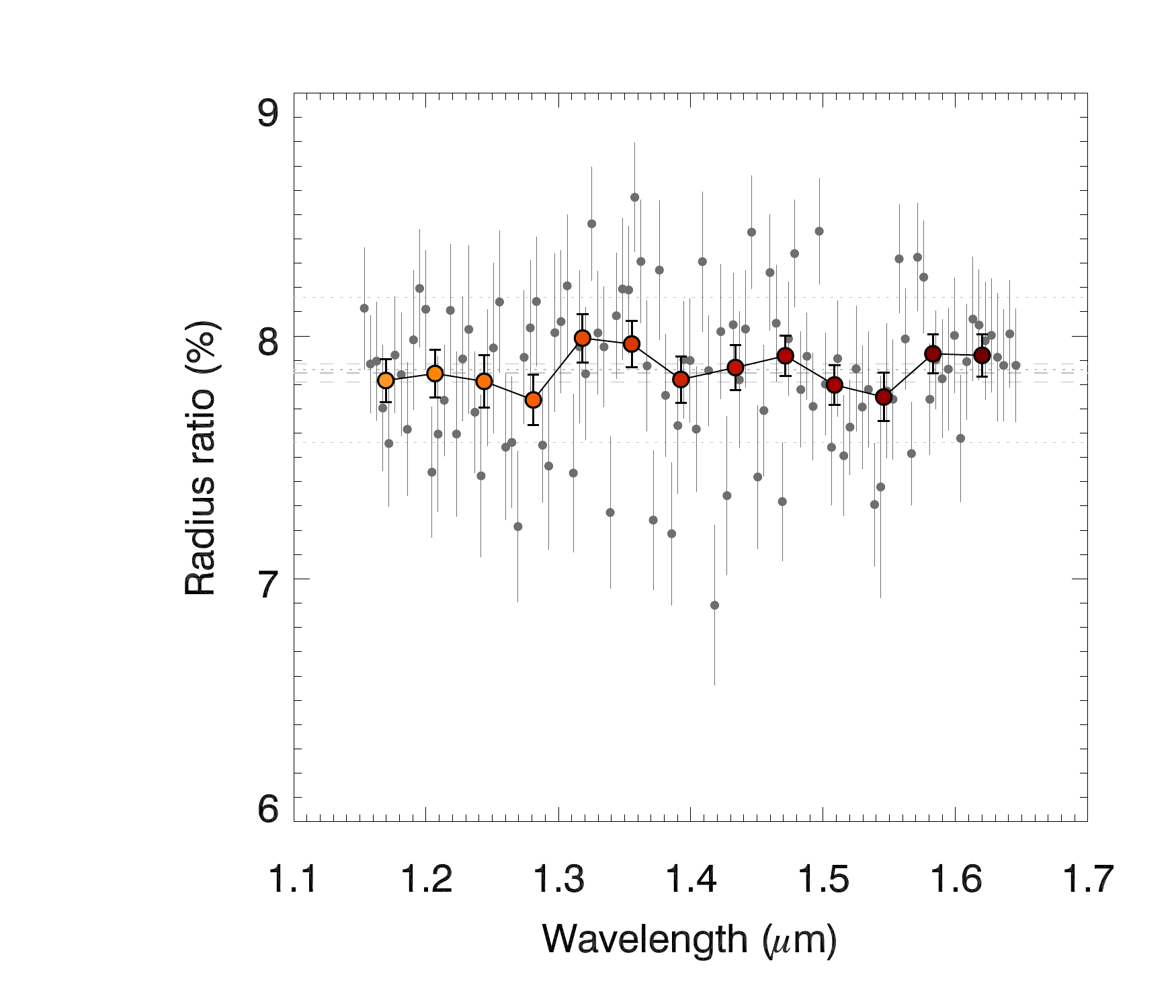}}
\resizebox{\columnwidth}{!}{\includegraphics[trim=2cm 0 0 1.3cm, clip=true]{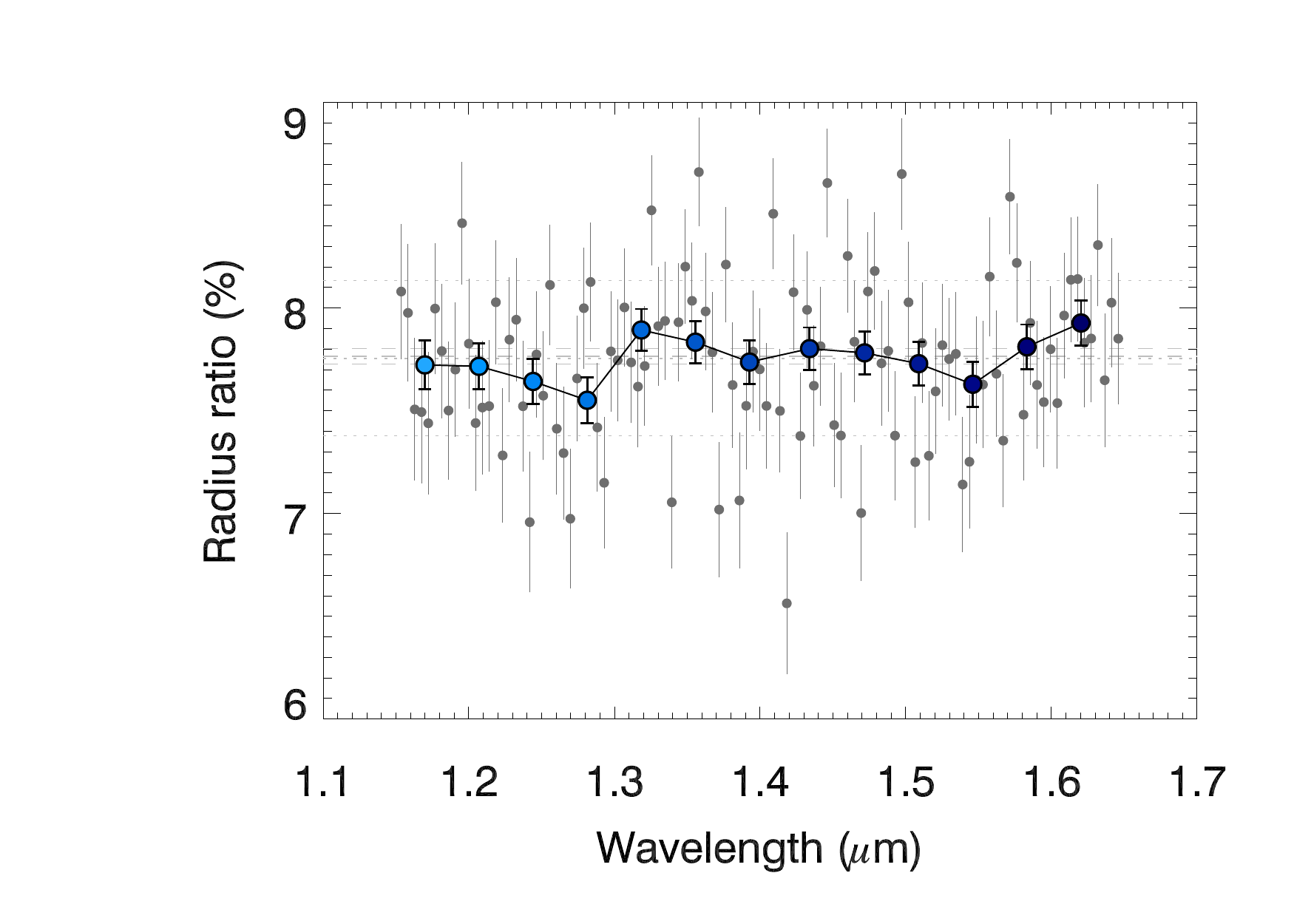}}
\caption{\label{fig:transpectrum-oot}WFC3/G141 transmission spectrum of GJ~3470b unbinned (grey dots) and binned by 8 (large coloured dots), obtained with the oot-divide \emph{(top panel)} and differential \emph{(bottom panel)} methods. The dotted horizontal lines indicate the mean and standard deviation values. The broad-band radius ratio from the white light curve and its 1-$\sigma$ uncertainty are shown by dashed horizontal lines.}
\end{figure}

The transmission spectrum of GJ~3470b results from (i) the radius ratios obtained by fitting each chromatic light curve or (ii) the differential flux measurements obtained from the white-light-transit corrected time series, depending on the method used (\texttt{divide-oot} or differential, respectively). Both methods yield similar results (cf.~Fig.~\ref{fig:transpectrum-oot} and Table~\ref{tab:radiusvalues}) and in the following we will focus on the results of the \texttt{divide-oot} procedure. The transit fits to all spectral channels are shown in the left panel of Fig.~\ref{fig:ootchromlc}, with their residuals in the right panel. The values of the radius ratio, reduced $\chi^2_\nu$, and rms of the residuals are listed as a function of wavelength in Table~\ref{tab:radiusvalues}. The mean rms of the residuals is 3659~ppm per $\sim 4.6$~nm wide spectral channel. 

\addtocounter{table}{1}

The WFC3/G141 transmission spectrum of GJ~3470b is plotted in Fig.~\ref{fig:transpectrum-oot} as the radius ratio variations against wavelength. At the level of uncertainty obtained on the chromatic radius ratio measurements (938~ppm for 37~nm wide spectral bins), the spectrum appears featureless. As can be seen in Fig.~\ref{fig:transpectrum-oot}, the broad-band radius ratio (the dashed horizontal line) provides a good straight-line approximation to the spectrum.

\section{Discussion}
\label{sec:discu}

\subsection{Comparison with other measurements}
\label{sec:allmeasurements}

\subsubsection{Review of measurements available in the literature}
Since the recent discovery of GJ~3470b, several measurements of its radius has been obtained from the near-ultraviolet to the infrared. All measurements reported in the literature are plotted in Fig.~\ref{fig:transpectrum-all-points}. 

A set of photometric measurements of GJ~3470b was first obtained by \citet{Fukui:2013hq}. These authors observed the transit in four broad bands (green points in Fig.~\ref{fig:transpectrum-all-points}) spread in the optical ($g'$, $R_c$, and $I_c$) and near-infrared ($J$), with the ISLE instrument on the 188~cm telescope of the Okayama Astrophysical Observatory ($J$ band) and the 50~cm MITSuME telescope (optical bands), set at the same location. The precision of these data increases towards the red. With the 188~cm telescope, \citet{Fukui:2013hq} obtained a good precision on the planet-to-star radius ratio, $(7.577^{+0.072}_{-0.075})$\% at $\sim1.2$~\micron, where a direct comparison with our data set is possible (see 
Sect.~\ref{sec:jband}). 

\citet{Crossfield:2013eu} measured the transit of GJ~3470b with the MOSFIRE multi-object spectrograph on the Keck~{\sc i} telescope. They obtained a transmission spectrum in the $K$ band, ranging from 2.09~\micron\ to 2.36~\micron\ (violet points in Fig.~\ref{fig:transpectrum-all-points}). These authors have also observed in the red optical with GMOS on Gemini-North; however, as they regard this latter data set as of poor quality (Crossfield et al.\ 2013 and I.~Crossfield, personal communication), we will not discuss the GMOS data. 

\citet{Nascimbeni:2013db} obtained transit measurements in two bands, one in the near-ultraviolet around 357.5~nm (the 37.5~nm wide $U_\mathrm{spec}$ band) and the other in a 20~nm band centred on 963.5~nm, thus flanking the whole optical domain. They observed simultaneously in these two bands with the LBC camera on the Large Binocular Telescope, and measured remarkably different planet-to-star radius ratios: $(8.21\pm0.13)$\% and $(7.484^{+0.052}_{-0.048})$\% in the UV and near-IR, respectively. These authors interpret this difference as the signature of a scattering process in the planetary atmosphere, which they claim should be cloud-free and with a low mean molecular weight.

Finally, as mentioned above, \citet{Demory:2013hv} measured the planet-to-star radius ratio to be $(7.798^{+0.046}_{-0.045}) \%$ in the 4.5~\micron\ channel of \emph{Spitzer}/IRAC.

\subsubsection{Comparison of $J$-band measurements}
\label{sec:jband}
\begin{figure*}
  \begin{minipage}[c]{0.65\textwidth}
    \includegraphics[width=\textwidth]{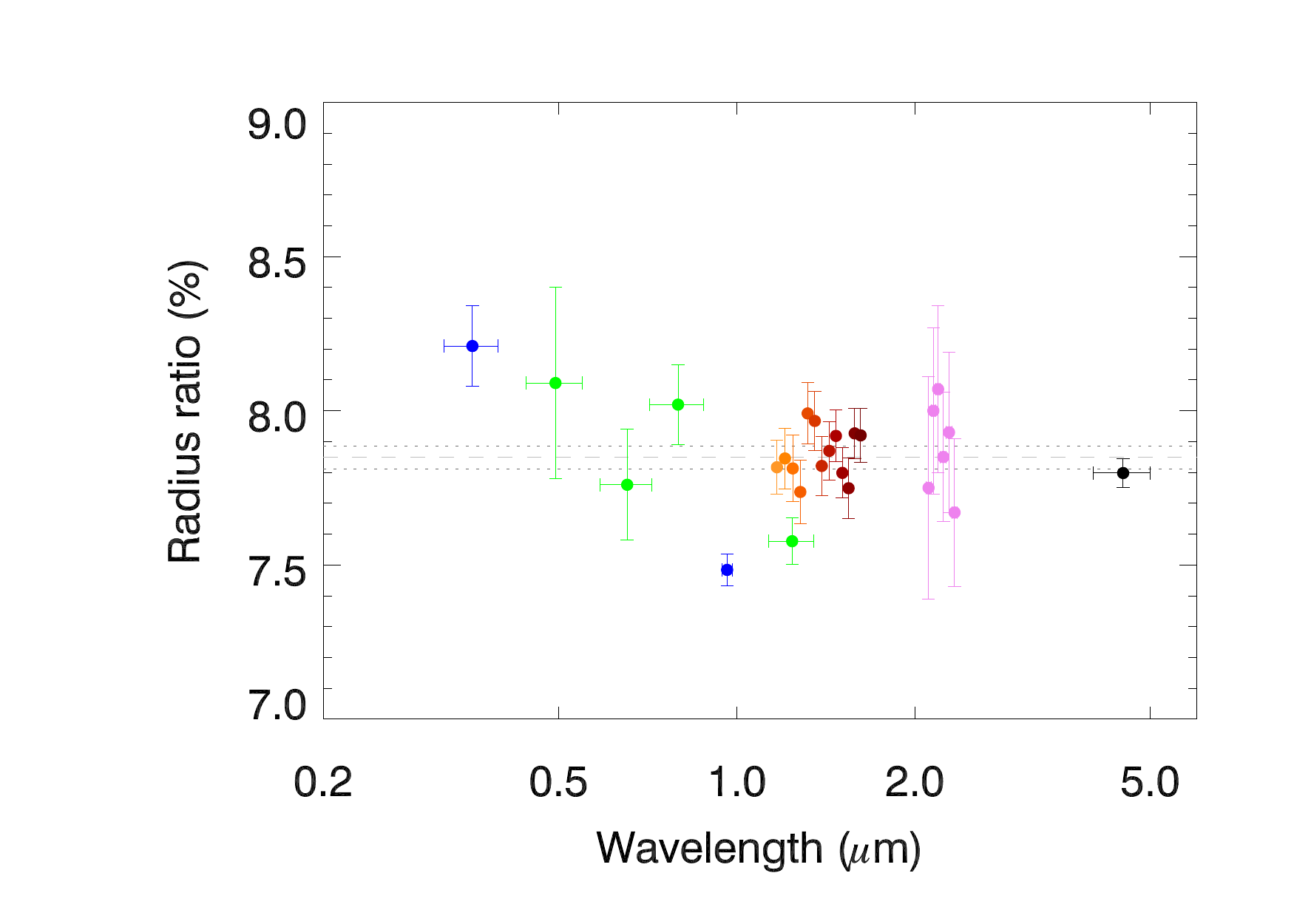}
  \end{minipage}\hfill
  \begin{minipage}[c]{0.35\textwidth}
    \caption{\label{fig:transpectrum-all-points}Transmission spectrum of GJ~3470b from the near ultraviolet to the infrared, with all published values reported: LBT measurements at 357.5 and 963.5~nm \citep[blue;][]{Nascimbeni:2013db}, OAO telescopes visible and near-infrared broad-band measurements \citep[green;][]{Fukui:2013hq}, near-infrared \hst/WFC3 spectroscopic measurements (orange to dark red; this work), Keck/MOSFIRE narrow-band measurements \citep[violet;][]{Crossfield:2013eu}, and \emph{Spitzer}/IRAC broad-band photometry at 4.5-\micron\ \citep[black;][]{Demory:2013hv}. The dashed and dotted lines show the white light curve radius ratio and its $\pm1$-$\sigma$ uncertainty, respectively.}
    \end{minipage}
\end{figure*}

As can be seen in Fig.~\ref{fig:transpectrum-all-points}, our \hst/WFC3 measurements lie above the $J$-band measurement of \citet{Fukui:2013hq} at $R_p/R_\star(J) = (7.577\pm0.072)\%$. Integrating the WFC3/G141 transmission spectrum over the $J$ band, we find a mean radius ratio of $(7.7837\pm0.0357)\%$, 
i.e. 3-$\sigma$ \citep[with the uncertainty of][]{Fukui:2013hq} to 6-$\sigma$ (with our uncertainty) higher than the ratio previously measured by \citet{Fukui:2013hq} in this band. We emphasise that our WFC3 spectrum yields an integrated value over the $J$ band which is similar to the value obtained over the $H$ band, on the red side of the $\varphi$ water band. As we will see in the next section, atmospheric models could not easily explain a significant difference between the $J$ and $H$ bands. 

Besides statistics, this $>3\sigma$ discrepancy could be due to unseen systematics in either data sets or astrophysical variations. This second possibility could in turn be the result of intrinsic stellar variations such as the presence of one or several (non-transited) stellar spots, which would have artificially increased the transit depth during the \hst\ observations. \citet{Fukui:2013hq} estimated the level of stellar variability needed to explain a difference in $R_p/R_\star$: the difference of $\Delta R_p/R_\star = (0.21\pm0.09)\%$ between our $J$-band measurement and that of \citet{Fukui:2013hq} could be explained by a $\sim5.5\%$ variability in the stellar flux. However, a two-month photometric survey of GJ~3470 in the $I_c$ band led \citet{Fukui:2013hq} to the conclusion that with a peak-to-peak variability of $\sim1\%$, the star should not be that active.

Another tantalising possibility is weather variations in the planetary atmosphere, in particular regarding the cloud coverage at the terminator: the observations of Fukui et al.\ (2013) could have been made under clear sky conditions at the limb of GJ~3470b (a scenario that these authors are actually proposing), whereas our observations would have taken place under covered conditions. To explain a 0.21\% difference in $R_p/R_\star$, the optically thick radius of GJ~3470b should change by $0.21\% R_\star \approx 1\,150$~km, or $\sim1.8$ atmospheric scale heights, which cannot be excluded. On the other hand, observing a similar behaviour for all points along the terminator seems a rather ad-hoc condition.

\subsubsection{Flat spectrum in the infrared}
While the other photometric measurements available cannot be directly compared to our data, they draw a dichotomic spectrum, where all measurements towards the red side of our data set are consistent with a flat spectrum, whereas the more scattered measurements towards the blue hint at an increasing slope typical of a light scattering process. The WFC3 spectrum thus plays a pivotal role in understanding the atmosphere of GJ~3470b.

The $K$-band radius ratio obtained by \citet{Crossfield:2013eu} is $(7.89^{+0.21}_{-0.19})$\%, at $0.2\sigma$ from our white radius ratio (using their error bars). Their binned spectrum is also flat within the error bars. In addition, the \emph{Spitzer}/IRAC broad-band photometric point at 4.5~\micron\ \citep{Demory:2013hv} is, as stated above, in agreement to $1.4\sigma$ with our broad-band measurement. Therefore, the \hst/WFC3, Keck/MOSFIRE, and \emph{Spitzer}/IRAC data are compatible with a flat transmission spectrum for GJ~3470b from $\sim1$ to 5~\micron. 

\subsection{Transmission spectrum model}
We will first focus on our \hst/WFC3 observations and attempt to determine whether it is possible to derive significant constraints on the atmospheric nature of GJ~3470b from this data set alone. Then, we will consider the full set of measurements described in the previous section, and see how the pivotal WFC3 data set can allow us to better understand the dichotomic nature of the transmission spectrum (slope increasing towards the blue below 1~\micron, flat spectrum above 1~\micron).

To these purposes, we compute theoretical transmission spectra for GJ~3470b, based on the radiative transfer model described in \citet{Ehrenreich:2006bm}. The code performs one-dimensional, single-scattering radiative transfer in grazing (limb) geometry across an atmospheric model, which structure and composition are user-specified. It calculates photo-absorption cross sections from atomic and molecular line lists \citep[all molecules present in the HITRAN data base;][]{Rothman:2009ea}, Rayleigh scattering from refractive indices for the main possible carriers (H$_2$, H$_2$O, CO$_2$, and N$_2$), and the effect of clouds and hazes. 

All the models discussed here are based on simulated atmospheres for GJ~3470b. The atmospheric structure is calculated under the assumptions of perfect gas mixtures in hydrostatic equilibrium with an isothermal temperature profile at 615~K. The code computes and integrates the opacities along the line of sight starting upward from a reference level set to 10~bar. The different models corresponds to different compositions of the gas mixture, expressed in volume mixing ratios (VMR, noted $X$), and various properties for clouds and hazes.

We distinguish between clouds and hazes along the following lines: \emph{clouds} are considered here as a uniform, optically thick layer with an achromatic behaviour. This means that only species absorbing or scattering light above the cloud layer can yield a spectroscopic signature in the otherwise flat transmission spectrum. On the other hand, \emph{hazes} are modelled as a vertical distribution of droplets (or grains) with characteristic sizes and size distribution widths, number densities, and altitude (or pressure) ranges. The different possible natures of haze (e.g., sulfuric acid, tholins, acetylene, etc.) are characterised by different wavelength-dependent complex refractive indices. Because tholins have been advanced as a possible solution to fit GJ~3470b's transmission spectrum \citep{Nascimbeni:2013db}, we will only consider \emph{tholin} haze in the following.

The full set of models and their parameters are listed in Table~\ref{tab:models}. We consider several families of models:
\begin{itemize}
\item{\emph{Cloud-covered atmosphere:}} An atmosphere covered by a high-altitude ($<10$~mbar) cloud layer. The transmission spectrum of this model is simply represented by a straight line. It remains flat whatever the atmospheric properties are beneath the cloud layer.
\item{\emph{Pure-Rayleigh atmospheres:}} Pure hydrogen (H$_2$) atmospheres, where the only spectral signature is due to Rayleigh scattering. These models can be clear (no clouds) or cloudy, with a cloud layer at different pressure levels (100~mbar and 10~mbar).  
\item{\emph{H$_2$-rich, metal-poor atmospheres:}} Clear or cloudy H$_2$-rich ($X_\mathrm{H_2} \sim 1$) atmospheres with small amounts of minor gases. The minor gases have a strong impact on the transmission spectrum because of their photo-absorption cross-sections: we mainly consider water vapour in various VMRs (1\%, 100~ppm, and 1~ppm). Cloud layers are also added at different pressure levels (100~mbar and 10~mbar). In one of these models, we add methane ($X_\mathrm{CH_4}=0.1\%$) to reproduce the thermal-equilibrium composition of a thick atmosphere with $\rm C/O < 1$ and $T \approx 600$~K \citep{Hu:2014gz}.
\item{\emph{H$_2$O-rich atmosphere:}} A pure water atmosphere (and no clouds).
\item{\emph{Hazy-H$_2$ atmospheres:}} A set of H$_2$-rich, metal-poor atmospheres with haze layers. The haze layers have different properties (particle size, particle density, and pressure range) and refractive indices corresponding to Titan tholins \citep{Khare:1984bj}. 
\end{itemize}

\begin{table*}[!t]
\caption{\label{tab:models}Atmospheric models computed in the present work.}
\begin{tabular}{cccccccccccc}
  \hline \hline
  Model                  & $X_\mathrm{H_2}$  & $X_\mathrm{H_2O}$ & $X_\mathrm{CH_4}$ & Clouds    & \multicolumn{3}{c}{Tholin hazes}       && \multicolumn{2}{c}{$\chi^2_\nu$} & Figs.\\
  \cline{6-8} \cline{10-11}
                              &                            &                              &                              &                & $r_m$       & $n$            & $p$            && WFC3 data    & All data         & \\
                              &                            &                              &                              & (mbar)     & (\micron) & (cm$^{-3}$) & (mbar)        && ($\nu=12$)  & ($\nu=24$)   & \\
  \hline
  Cloud-covered    & --                       & --                         & --                         & $<10$     & --           & --              & --              && 8.53              & 67.2              & \ref{fig:transpectrum-wfc3-models}, \ref{fig:transpectrum-full-models-simple}\\
  \hline
  Pure Rayleigh      & 1                         & 0                            & 0                           & --           & --            &  --              & --             && 12.3              & 92.7              & \ref{fig:transpectrum-full-models-simple}\\
                              & 1                        & 0                            & 0                           & 100         & --            & --               & --             && 8.59              & 24.9             & \ref{fig:transpectrum-full-models-simple}\\
                              & 1                        & 0                            & 0                           & 10           & --            & --               & --             && 8.51              & 11.6             & \ref{fig:transpectrum-full-models-simple}\\
  \hline
 H$_2$-rich           & 1                         & 1\%                        & 0.1\%                    & --           & --            &  --              & --             && 237            & 352             & \ref{fig:transpectrum-wfc3-models}, \ref{fig:transpectrum-full-models-h2o}\\
                              & 1                         & 1\%                       & 0                          & --           & --            &  --              & --              && 108              & 213              & \ref{fig:transpectrum-wfc3-models}, \ref{fig:transpectrum-full-models-h2o}\\
                              & 1                         & 100 ppm               & 0                          & --           & --            &  --              & --              && 58.6              & 147              & \ref{fig:transpectrum-wfc3-models}, \ref{fig:transpectrum-full-models-h2o}\\
                              & 1                        & 100 ppm                & 0                          & 100         & --           & --               & --              && 12.4               & 69.2             & \ref{fig:transpectrum-wfc3-models}, \ref{fig:transpectrum-full-models-h2o}\\
                              & 1                        & 100 ppm                & 0                          & 10           & --           & --               & --              && 9.27               & 66.1             & \ref{fig:transpectrum-wfc3-models}, \ref{fig:transpectrum-full-models-h2o}\\
                              & 1                        & 1 ppm                    & 0                          & 100         & --           &  --              & --              && 8.26               & 59.4            & \ref{fig:transpectrum-full-models-h2o}\\
  \hline
  H$_2$O-rich       & 100 ppm             & 1                            & 0                          & --           & --            & --              & --              && 8.79               & 73.1               & \ref{fig:transpectrum-wfc3-models}, \ref{fig:transpectrum-full-models-h2o}\\
  \hline
  Hazy H$_2$        & 1                          & 100 ppm                & 0                          & --           & 0.1          &  1000         & 100--1       && 26.0              & 135               & \ref{fig:transpectrum-full-models-hazy}\\
                             & 1                         & 100 ppm                & 0                          & --           & 0.1          &  1000         & 0.1--0.001 && 33.8              &  218              & --\\
                             & 1                         & 100 ppm                & 0                          & --           & 0.1          &   100          & 100--1       && 59.6              & 318               & \ref{fig:transpectrum-wfc3-models}, \ref{fig:transpectrum-full-models-hazy}\\
                             & 1                         & 100 ppm                & 0                          & --           & 0.1          &    100         & 0.1--0.001 && 126              & 943              & --\\
                             & 1                         & 100 ppm                & 0                          & --           & 0.1          &    10           & 100--1       && 51.9              & 199              & \ref{fig:transpectrum-full-models-hazy}\\
                             & 1                         & 100 ppm                & 0                          & --           & 0.1          &    10           & 0.1--0.001 && 56.9              & 468              & --\\
                             & 1                         & 100 ppm                & 0                          & --           & 0.5          &  10             & 100--1       && 24.9              & 99.6              & \ref{fig:transpectrum-full-models-hazy}\\
                             & 1                         & 100 ppm                & 0                          & --           & 0.5          &  10             & 0.1--0.001 && 33.8             & 116              & \ref{fig:transpectrum-full-models-hazy}\\
  \hline
\end{tabular}
\end{table*}

As increasing evidence of flat exoplanet transmission spectra are unveiled \citep[e.g.,][]{Kreidberg:2014du,Knutson:2014hz}, it becomes more and more obvious that scattering effects should be dealt with when modelling these transmission spectra. Mie scattering by hazes has often been advanced as an efficient way to flatten a transmission spectrum, and it sometimes does as in the case the acid sulfuric haze on Venus \citep{Ehrenreich:2012iq}. In this work, we have considered a haze of tholins instead of sulfuric acid droplets, because it is the haze model stressed by \citep[][see their Fig.~2c]{Nascimbeni:2013db} as providing the best fit to the radius ratio measurements of GJ~3470b known at the time of their publication. 

\citet{Nascimbeni:2013db} rescaled transmission spectrum models calculated by \citet{Howe:2012et} for a 10~\Mearth\ planet with an equilibrium temperature of 700~K (Howe \& Burrows' closest grid point to GJ~3470b; V.~Nascimbeni, personal communication). Because the atmospheric scale height scales linearly with temperature and is inversely proportional to surface gravity and mean molar mass of the atmosphere, it is certainly fine to rescale a transmission spectrum model to a planet with a different set of such parameters, as long as we can consider the atmosphere is in hydrostatic equilibrium and the VMRs of the different absorbing or (Rayleigh) scattering gases are constant. \emph{We caution, however, that this may not be appropriate when dealing with droplet or particle hazes}. In fact, the vertical structure of the haze layer (number density, pressure range) and the intrinsic parameters of the haze particles (size and size distribution width) are not following hydrostatic equilibrium. In other words, the haze properties driving the haze scattering signature, such as the particle size and number density, cannot be scaled with the density or partial pressure $p_i$ of gases, which follow the hydrostatic law $p_i = p_\mathrm{ref}\exp{(-z/H)}$, where $p_\mathrm{ref}$ is the pressure at the reference level, $z$ is the altitude, and $H$ is the atmospheric scale height. Instead of rescaling models existing for other planets, we have computed the scattering signature of haze layers for the set of haze parameters detailed in Table~\ref{tab:models}. 

\subsubsection{Interpretation of the \hst/WFC3 data}
\label{sec:modelwfc3}
\begin{figure*}
\begin{minipage}[c]{0.65\textwidth}
\includegraphics[width=\textwidth]{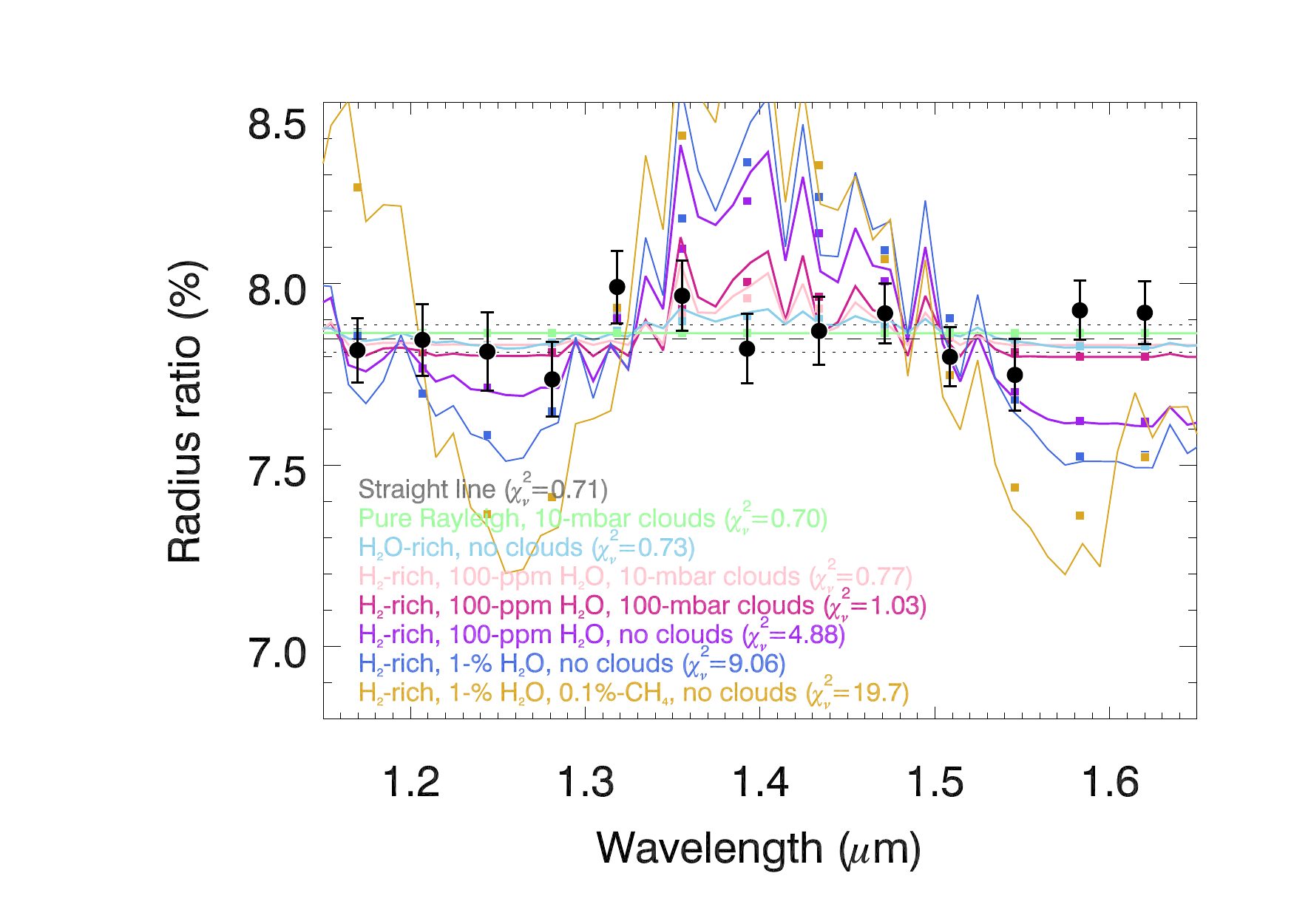}
\end{minipage}\hfill
\begin{minipage}[c]{0.35\textwidth}
\caption{\label{fig:transpectrum-wfc3-models}WFC3/G141 transmission spectrum of GJ~3470b (same as in Fig.~\ref{fig:transpectrum-oot}) vs. several transmission spectrum models. The white light curve radius ratio and its $\pm1$-$\sigma$ uncertainty are represented by the horizontal dashed and dotted lines, respectively.}
\end{minipage}
\end{figure*}

The modelled transmission spectra are over-plotted on the WFC3 data in Fig.~\ref{fig:transpectrum-wfc3-models}. The different models are adjusted to the data using a least-square approach where the only free parameter is an offset in $R_p/R_\star$. The  $\chi^2$ values are reported in the penultimate column of Table~\ref{tab:models}.

Our data are inconsistent with all models predicting a strong contrast between the $J$ and $H$ bands on the one hand, and the region separating those two bands on the other hand (where the 1.38~\micron\ water band is). In such models, the contrast is due to the photo-absorption of water, which effect is favoured in atmospheres with low mean molar masses, i.e., with large scale heights. Extra absorbers, such as methane, can also increase the absorption if, at a given wavelength, they have larger cross sections than water (and are located as high as water or above in the atmosphere).

Among all the tested models, the H$_2$-rich model with 1\% of H$_2$O and 0.1~\% of CH$_4$ is the most realistic in terms of composition because it follows the predictions of \citet[][see their Fig.~7, upper left panel]{Hu:2014gz} for a $\sim 600$-K planet with a thick, hydrogen-rich, $\rm C/O<1$ atmosphere in thermal equilibrium. This model also predicts the largest contrast in the WFC3 band; it yields a $\chi^2$ of 237 for $\nu=12$ degrees of freedom, and is thus inconsistent with the WFC3 data to the $\sqrt{\chi^2}=15$-$\sigma$ level.

All other H$_2$-rich models are methane-free, which reduces the predicted contrast with the $J$ and $H$ bands. However, the observed transmission spectrum is sufficiently flat to exclude a cloudless, H$_2$-rich atmosphere with 1\% of water ($\chi^2=108$, $\nu=12$) with a good confidence level (10$\sigma$). We found that lowering the thermochemistry equilibrium VMR of water by two orders of magnitudes (to $X_\mathrm{H_2O} = 100$~ppm) does not sufficiently decrease the amplitude of the water band. This latter model ($\chi^2=58.6$, $\nu=12$) is also significantly (8$\sigma$)  excluded by our data.

Only models yielding less contrasted transmission spectra can provide good fits to the data. There are mainly two ways to flatten the transmission spectrum: (i) adding cloud layers and (ii) significantly increasing the mean molar mass of the atmosphere, by turning minor species into major components.

(i) An efficient way to flatten the transmission spectrum of H$_2$-rich atmospheres is to introduce a cloud layer. The quality of the fit with the H$_2$-rich, 100~ppm H$_2$O model significantly improves for a cloud layer set at 100~mbar ($\sim$2\,000~km above the 10~bar level; $\chi^2=12.4$ for $\nu=12$) and 10~mbar ($\sim$3\,000~km above the 10~bar level; $\chi^2=9.3$ for $\nu=12$). Setting the cloud top at even lower pressures will eventually suppress the water band; however it is unknown whether optically-thick clouds could exist at these low-pressure levels. A model with very high clouds would produce a perfectly flat transmission spectrum. The best-fit straight-line model yields a $\chi^2=8.5$ for $\nu=12$). We emphasise that the atmosphere could have any possible compositions below such a high-altitude cloud, and we would not be able to discriminate between any of them. 

We found that contrary to clouds, tholin haze layers are unable to flatten the transmission spectrum in the WFC3 band pass. The best (flattest) results, with $\chi^2\approx25$ for $\nu=12$, are obtained for large particle sizes (0.5~\micron) or high particle density (1\,000~cm$^{-3}$) located at relatively high pressures (between 100~mbar and 1~mbar) in the atmosphere.

(ii) The cloudless water-rich model, where the VMR of water is set to 99.99\%, also provides a good fit to the WFC3/G141 data ($\chi^2=8.8$ for $\nu=12$). This is because the atmospheric scale height in this model is $\rm \mu_{H_2O} / \mu_{H_2} = 9$ times smaller than in the H$_2$-rich models, efficiently decreasing the amplitude of the absorption features. We note, however, that according to \citet{Hu:2014gz}, an almost 100\%-water atmosphere is unlikely, as large H$_2$O VMRs ($X_{\rm H_2O}\ga10\%$) also require larger hydrogen VMRs ($X_{\rm H_2}\ga10\%$) than the value we have considered ($X_{\rm H_2}=0.01\%$). For very low $X_{\rm H_2}$, the atmosphere would in fact be dominated by $\rm O_2$ or $\rm CO_2$ ($\rm C/O \la 0.5$) or by $\rm CH_4$ ($\rm C/O \sim 1$) at the considered temperature \citep[][see their Fig.~7]{Hu:2014gz}. The net effect of having more hydrogen in a water-rich atmosphere is to decrease the mean molar mass, and thus to increase the scale height and the expected water absorption feature. This could make a water-rich model more inconsistent with our data. On the other hand, a hydrogen-poor atmosphere dominated, as stated above, by gases heavier than water vapour (O$_2$, CO$_2$) or having equivalent molar mass (CH$_4$), will have an atmospheric scale height lower or equal, respectively, to our water-rich model. Therefore, such atmospheric compositions are expected to yield even more flat transmission spectra in the wavelength range covered by our WFC3 measurements. Hence, CO$_2$-rich or O$_2$-rich atmospheres would also yield transmission spectra fully consistent with the WFC3/G141 spectrum.

\subsubsection{Interpretation of the full transmission spectrum}

\begin{figure*}
\begin{minipage}[c]{0.65\textwidth}
\includegraphics[width=\textwidth]{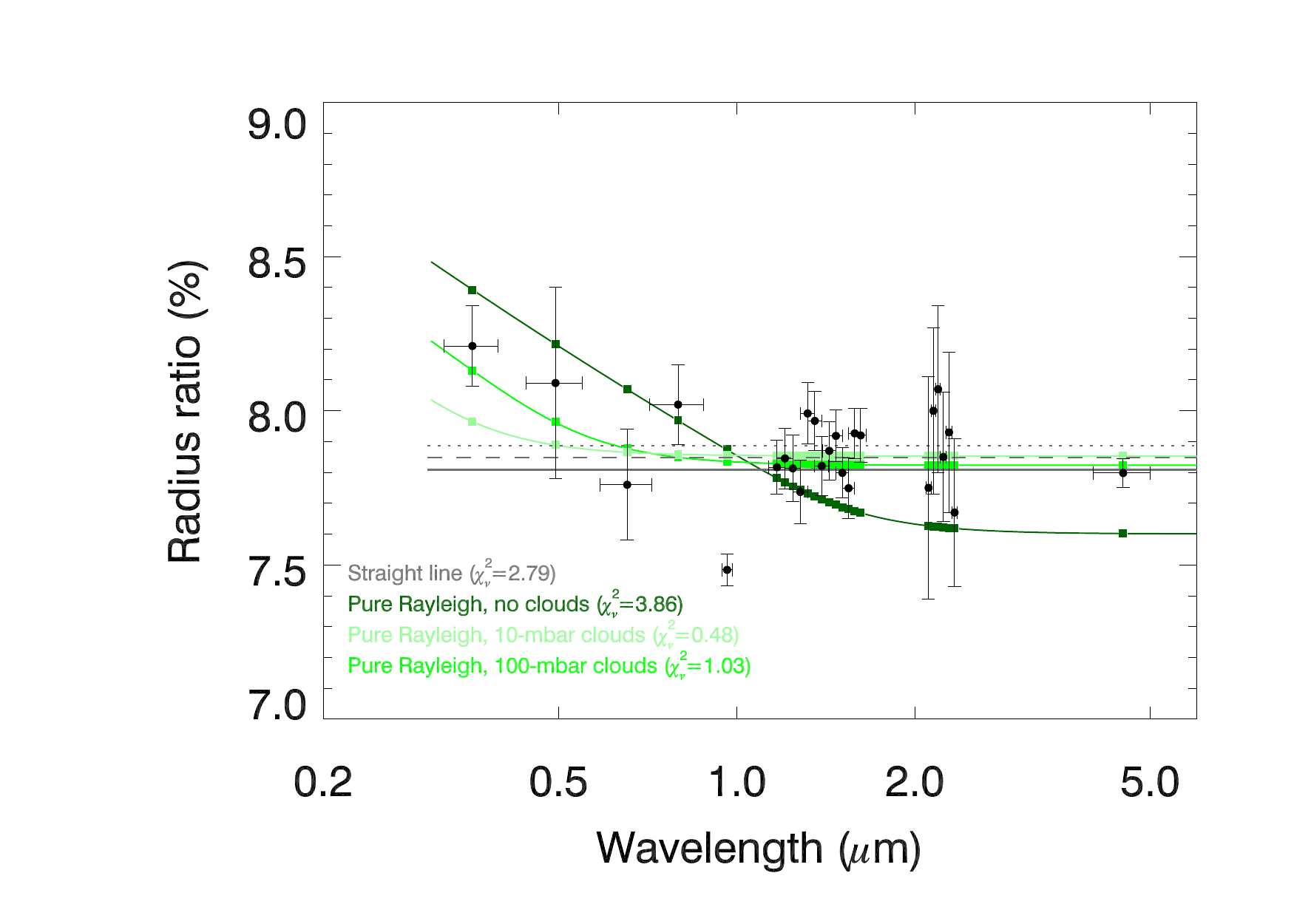}
\end{minipage}\hfill
\begin{minipage}[c]{0.35\textwidth}
\caption{\label{fig:transpectrum-full-models-simple} Simple transmission spectrum models: straight-line (thick grey horizontal line) and pure-Rayleigh's (green curves) vs. all radius ratio measurements of GJ~3470b (black dots; same as in Fig.~\ref{fig:transpectrum-all-points}). Squares show the models value over the measurement band passes. The WFC3 broad-band radius ratio and its 1-$\sigma$ uncertainty are represented by the dashed and dotted lines, respectively.}
\end{minipage}
\end{figure*}

\begin{figure*}
\begin{minipage}[c]{0.65\textwidth}
\includegraphics[width=\textwidth]{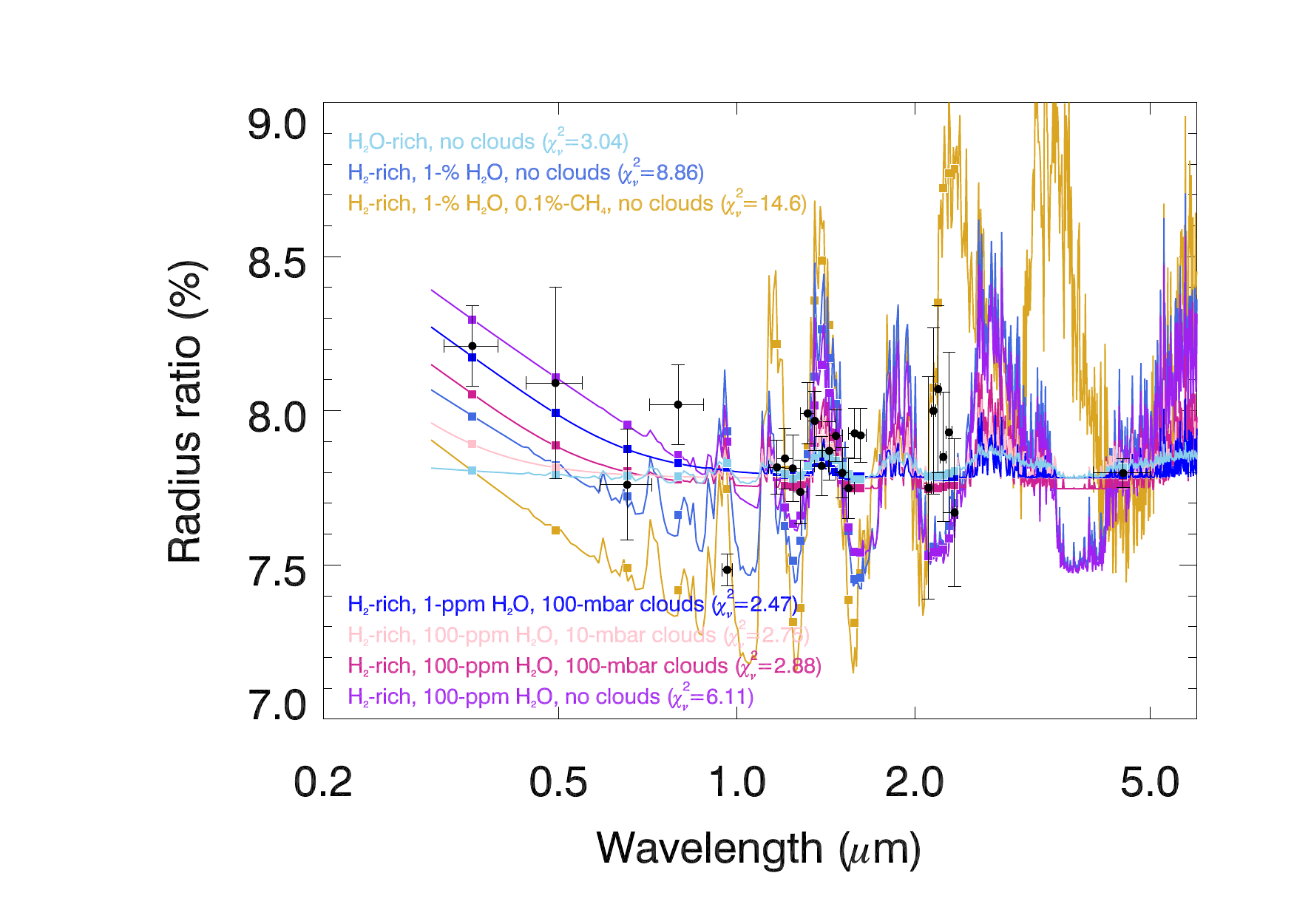}
\end{minipage}\hfill
\begin{minipage}[c]{0.35\textwidth}
\caption{\label{fig:transpectrum-full-models-h2o} Same as Fig.~\ref{fig:transpectrum-full-models-simple} with water-rich and water-poor (with or without clouds) transmission spectrum models.}
\end{minipage}
\end{figure*}

\begin{figure*}
\begin{minipage}[c]{0.65\textwidth}
\includegraphics[width=\textwidth]{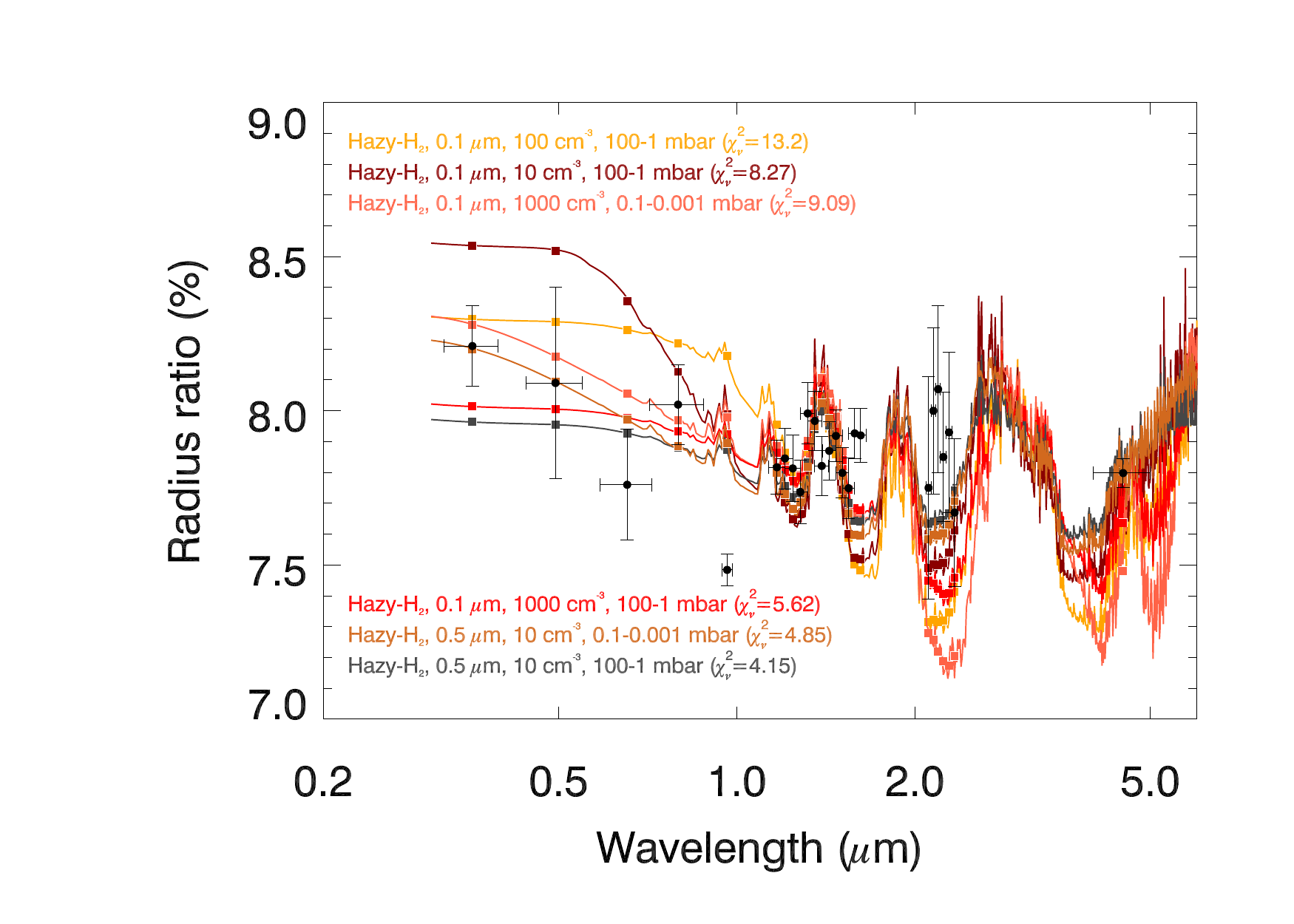}
\end{minipage}\hfill
\begin{minipage}[c]{0.35\textwidth}
\caption{\label{fig:transpectrum-full-models-hazy} Same as Fig.~\ref{fig:transpectrum-full-models-simple}  with hazy hydrogen-rich models.}
\end{minipage}
\end{figure*}

We now consider the full set of $R_p/R_\star$ measurements for GJ~3470b (Fig.~\ref{fig:transpectrum-all-points}) and compare it with the models discussed above. We again fit the data by adjusting a vertical offset to the different models. Figures~\ref{fig:transpectrum-full-models-simple} to~\ref{fig:transpectrum-full-models-hazy} show a comparison of all these data with our models. The idealistic models (cloud-covered and pure-Rayleigh) are shown in Fig.~\ref{fig:transpectrum-full-models-simple}. Figure~\ref{fig:transpectrum-full-models-h2o} shows the different H$_2$-rich models with minor species and the H$_2$O-rich model, and Fig.~\ref{fig:transpectrum-full-models-hazy} shows the hazy-H$_2$ models. 

As detailed above in Sect.~\ref{sec:allmeasurements}, the data are consistent with a flat transmission spectrum between 1~\micron\ and 5~\micron, whereas there is a trend for a increasing slope towards short wavelengths below 1~\micron. 

All models predicting non-flat infrared spectra are inconsistent with the data collected beyond 1~\micron. This excludes the models with tholin hazes (Fig.~\ref{fig:transpectrum-full-models-hazy}), which do not hide the water or methane molecular signatures and thus yield $\chi^2 \geq 100$ for $\nu=24$ (see Table~\ref{tab:models}). Cloudless H$_2$-rich models with minor species (Fig.~\ref{fig:transpectrum-full-models-h2o}) are also ruled out: these models predict strong water and/or methane signatures in the infrared, yielding $147<\chi^2<352$ for $\nu=24$. Adding clouds to the H$_2$-rich, 100~ppm H$_2$O model helps in obtaining better fits ($\chi^2=69.2$ and 66.1 for $\nu=24$, for cloud top at 100~mbar and 10~mbar, respectively), as this reduces the amplitude of the molecular signatures. For these models, Rayleigh scattering towards short wavelengths yields a slope similar to that suggested by data collected in the visible. These models, however, are not flat enough beyond 1~\micron\ to satisfyingly match the infrared data.

On the other hand, the models tested above in Sect.~\ref{sec:modelwfc3} to yield a flat infrared spectrum, consistently yield a flat spectrum in the visible. Increasing the VMR of minor species (water or methane, both with a molar mass of 18~g~mol$^{-1}$) would eventually lead to shrink the atmospheric scale height and compress all spectroscopic signatures, including the Rayleigh scattering slope towards short wavelengths. This is why the H$_2$O-rich model (sky blue curve in Fig.~\ref{fig:transpectrum-full-models-h2o}), which provided a good fit to the WFC3 data only ($\chi^2=8.8$ for $\nu=12$), does not work as well when the full set of data is considered ($\chi^2=73.1$ for $\nu=24$). Similarly, the cloud-covered model producing a straight-line transmission spectrum is not favoured either when all the data are considered ($\chi^2=67.2$ for $\nu=24$), because of the increasing slope towards short wavelength suggested by the optical measurements. With the achieved level of precision, it is fair to say that we cannot distinguish between a cloudy H$_2$-rich atmosphere with $X_\mathrm{H_2O}=100$~ppm, a pure water atmosphere, or a cloud-covered atmosphere yielding a completely flat transmission spectrum. All these models yield $\chi^2 \approx 70$ ($\nu=24$). Consequently, none of these models satisfyingly fit both visible and infrared data.

Obtaining a better match between models and observations would be achieved by an atmospheric model producing a featureless spectrum in the infrared while preserving a Rayleigh slope in the visible. One possibility, that we are now going to explore, would be to further reduce the VMR of H$_2$O in the H$_2$-rich atmospheres with minor gases and clouds. The decrease in VMR would have to be significant in order to sufficiently attenuate the molecular signatures, because of the analytical degeneracy between the abundance of a species and the total pressure\citep{LecavelierdesEtangs:2008he}: we have tested the case of a cloudy model (clouds at 100~mbar) with an extremely low VMR of water (1~ppm), for which we indeed obtain a decreased $\chi^2$ of 59.4 ($\nu=24$), with respect to the previously tested model with $X_\mathrm{H_2O} = 100$~ppm and clouds at the same altitude (see Fig.~\ref{fig:transpectrum-full-models-h2o}).

Such H$_2$-rich atmospheres with extremely low or no amount of water are what we call the pure-Rayleigh H$_2$ atmospheres. These ideal models, plotted in Fig.~\ref{fig:transpectrum-full-models-simple}, actually yield the best fits to the whole data set because they are flat beyond 1~\micron\ while featuring a Rayleigh slope in the visible. The amplitude of the slope depends on the presence and altitude of clouds. For cloud layers at 100 and 10~mbar, pure-Rayleigh H$_2$ models yield $\chi^2$ of 24.9 and 11.6, respectively (for $\nu=24$), which is a significant improvement with respect to the previously tested models. 

While pure-Rayleigh H$_2$ atmospheres represent a good match to the available data, they are not necessarily physically or chemically realistic. In fact, having no or extremely low ($<1$~ppm) amounts of water in a H$_2$-rich ($X_\mathrm{H_2} > 0.9$) atmosphere is at odds (by $\sim$4 orders of magnitude) with the predictions of the photo- and thermochemical model of \citet{Hu:2014gz}. According to these authors, the water VMR could be lower than 1~ppm if $X_\mathrm{H_2} \la 0.7$ and $\rm C/O > 2$. This would result in a significantly increased amount of methane ($X_\mathrm{CH_4} \ga 10\%$), involving strong spectroscopic signatures in the infrared, inconsistent with the data, as the H$_2$-rich model with $X_\mathrm{CH_4}=0.1\%$ suggests (golden curve in Fig.~\ref{fig:transpectrum-full-models-h2o}). The \citet{Hu:2014gz} carbon-rich models also predict large amounts of hydrocarbon species, such as ethylene ($\rm C_2H_4$) and acetylene ($\rm C_2H_2$), which may produce hazes with refraction indices different from the tholin haze we have tested. The effect of such hazes on the transmission spectrum remains to be evaluated, even though the transmission signature of a haze of acetylene has been tested by \citet{Howe:2012et} for GJ~1214b. These authors concluded that `for polyacetylene particles of any size, the wavelength dependence is weak at short wavelengths', suggesting that such a haze cannot account for the tentative slope observed on GJ~3470b transmission spectrum.

One way to understand a significant depletion of water deduced from a transmission spectrum such as the one obtained for GJ~3470b (scattering slope in the visible, flat in the infrared) is to consider that such spectra only probe the terminator regions of the planet. We speculate that if the heat redistribution in the planetary atmosphere is not efficient (meaning that the atmospheric advection timescale is lower than the radiative timescale), then the terminator would represent the boundary between the hot illuminated side of the planet and the cold night side. As such, the evening terminator could be the place where water condenses and snow flakes settle down to pressure ranges below the optically thick level. Water vapour VMR could increase again to the predicted values after the atmospheric circulation brings it passed the morning terminator. If the visible slope of GJ~3470b spectrum is observationally confirmed, this hypothesis could be tested with three-dimensional, global circulation models.

\section{Conclusion}
In this work, we have presented the first space-borne transmission spectrum of GJ~3470b, a warm, Uranus-mass exoplanet. The near-infrared spectroscopic measurements of the WFC3/G141 extend from 1.1~\micron\ to 1.7~\micron, covering the $J$ and $H$ photometric band passes as well as the $1.38$~\micron\ water absorption separating them. The WFC3/G141 data, obtained in stare mode, offer a planet-to-star radius ratio precision of $\sim0.09\%$ per $\sim40$-nm bin. The data show no sign of the water band at 1.38~\micron\ and are significantly inconsistent ($>10\sigma$) with models where it is prominent, such as a cloudless H$_2$-dominated atmosphere. Our results are indeed consistent, within the error bars, with a featureless spectrum. 

Our simulations of theoretical transmission spectra for GJ~3470b show that the \hst/WFC3 data could be compatible with: (i) a cloudy hydrogen-rich atmosphere with trace amounts of water ($<100$~ppm), (ii) a pure water atmosphere, or (iii) a cloud-covered atmosphere, with high-altitude optically thick clouds ($<10$~mbar). This leads us to advance that the transmission spectrum of GJ~3470b is flat in the infrared, akin to what is seen -- with a higher precision -- for the warm super-earth GJ~1214b \citep{Kreidberg:2014du} and the warm neptune GJ~436b \citep{Knutson:2014hz}. 

Meanwhile, GJ~3470b stands out with respect to these two exoplanets because the full set of its available measurements, from the near-ultraviolet to the infrared, points to a dichotomic transmission spectrum: flat beyond 1~\micron\ (as demonstrated by the present work) but with a tentative slope rising towards shorter wavelengths in the optical.\footnote{After this paper was refereed, we became aware of the study by \citep{Biddle:2014bx}, who have re-analysed the published ground-based measurements. This work add credence to the Rayleigh slope, which fully supports our conclusions.} We ruled out tholin haze as a good candidate to explain this spectroscopic behaviour. Instead, we found that the only atmospheric model able to satisfyingly match the data is a cloudy hydrogen-rich atmosphere with an extremely low water VMR ($<1$~ppm) or so-called pure Rayleigh model. Such an atmosphere, however, challenges the one-dimensional thermo- and photo-chemical models \citep[e.g.,][]{Hu:2014gz} predicting much higher water VMRs ($\sim1\%$). 

An optical transmission spectrum of GJ~3470b would be extremely valuable in order to assess the reality of the tentative spectral slope, and thus the possibility that the atmosphere of this exoplanet is water-depleted at the terminator. On the other hand, if the spectral slope is ruled out by new observations in the optical, then increasing the precision on the near-infrared measurements, by observing additional transits with a much higher duty cycle (as made possible by the WFC3 scan mode), would allow discriminating between a cloudy atmosphere and a water-rich atmosphere, similarly to what has been achieved for GJ~1214b and GJ~436b \citep{Kreidberg:2014du,Knutson:2014hz}.

\acknowledgement
D.E. is grateful to T.-O.~Husser for his help with the use of the stellar atmosphere library, I.~Crossfield and T.~Barman for discussions related to the transmission spectrum of the planet. The authors would like to thank the anonymous referee for his or her time spent reviewing the manuscript. This work has been carried out within the frame of the National Centre for Competence in Research `PlanetS' supported by the Swiss National Science Foundation (SNSF). D.E., C.L., S.U., and D.S. acknowledge the financial support of the SNSF. X.B. acknowledges funding from the European Research Council under the ERC Grant Agreement n.~337591-ExTrA and from the French Agence National de la Recherche (ANR) under programme ANR-12-BS05-0012 `Exo-Atmos'. N.C.S. aknowledges the support from the European Research Council/European Community under the FP7 through Starting Grant agreement number 239953. N.C.S. was also supported by Funda\c{c}\~ao para a Ci\^encia e a Tecnologia (FCT) through the Investigador FCT contract reference IF/00169/2012 and POPH/FSE (EC) by FEDER funding through the program `Programa Operacional de Factores de Competitividade - COMPETE'. Finally, the authors would like to thank the Space Telescope Science Institute Director M.~Mountain for awarding Director's Discretionary Time to this project, and the programme coordinator Tricia Royle for, as always, a flawless implementation of the observations.

\bibliographystyle{aa}
\bibliography{23809.bib} 

\appendix

\section{Chromatic transit light curves}

\begin{figure*}[!p]
\resizebox{0.7\textwidth}{!}{\includegraphics{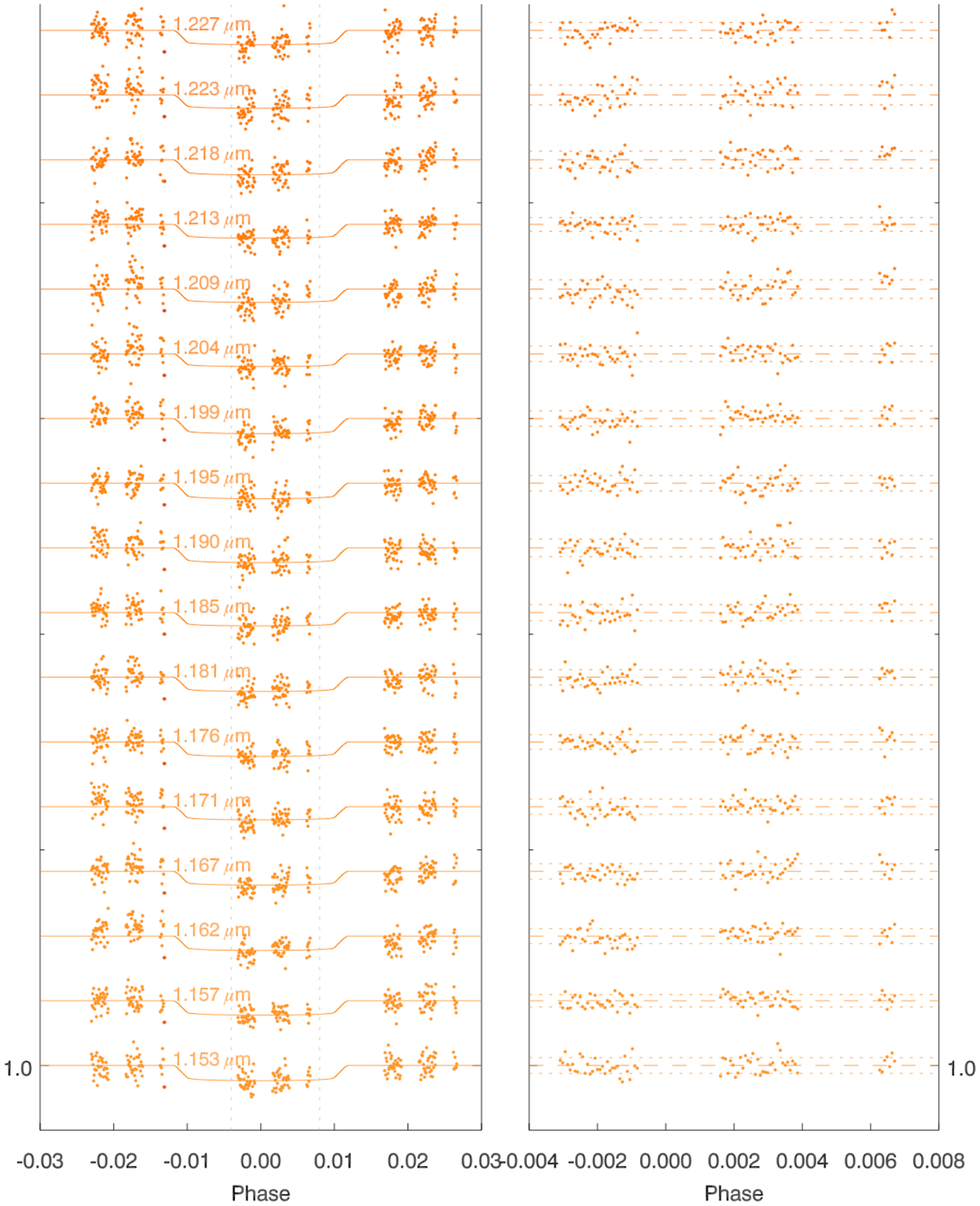}}
\caption{\label{fig:ootchromlc}The \emph{left panel} shows the 107 corrected chromatic light curves of GJ~3470. Data points in all three \hst\ orbits shown (2, 3, and~4) have been divided by the correction template. The light curves have been offset in flux for clarity. The plain curves are best-fit transit models superimposed to each time series. The grey vertical dotted lines indicate the phase coverage shown in the \emph{right panel}, where the residuals are plotted. In this panel, the horizontal dashed lines show the zero levels and the dotted lines show the one-standard-deviation levels of the residuals for the in-transit (third) \hst\ orbit only. \emph{Because of file size constraints on arXiv, only one panel of this figure is reproduced here. The complete figure is available upon request.}} 
\end{figure*}

\section{Transmission spectrum data}

\longtab{1}{
\begin{flushleft}
\begin{longtable}{lccccc}
\caption{\label{tab:radiusvalues}Transmission spectrum of GJ~3470b.}\\
\hline\hline
                     & \multicolumn{3}{c}{Out-of-transit data correction (Sect.~\ref{sec:oot-divide})} && Differential correction (Sect.~\ref{sec:differential-correction})\\
\cline{2-4}  \cline{6-6}
  Wavelength & $R_p/R_\star$ (\%)        & Light curve fit                     & Flux residual && $R_p/R_\star$ (\%)\\
  (\micron)    & [$\sigma_{R_p/R_\star}$] & $\chi^2_\nu$  ($\nu = 70$) & rms (ppm)     && [$\sigma_{R_p/R_\star}$] \\
\hline
\endhead
1.153 & 8.114 [0.251] & 0.92 & 3652 && 8.080 [0.330] \\
1.157 & 7.884 [0.202] & 0.56 & 2850 && 7.976 [0.334] \\
1.162 & 7.896 [0.244] & 0.86 & 3444 && 7.506 [0.346] \\
1.167 & 7.703 [0.260] & 0.93 & 3575 && 7.493 [0.345] \\
1.171 & 7.557 [0.258] & 0.91 & 3492 && 7.439 [0.345] \\
1.176 & 7.921 [0.240] & 0.87 & 3402 && 7.997 [0.319] \\
1.181 & 7.841 [0.254] & 0.96 & 3556 && 7.790 [0.326] \\
1.185 & 7.615 [0.275] & 1.08 & 3742 && 7.501 [0.336] \\
1.190 & 7.984 [0.287] & 1.31 & 4091 && 7.700 [0.326] \\
1.195 & 8.195 [0.243] & 1.00 & 3555 && 8.412 [0.296] \\
1.199 & 8.110 [0.244] & 1.01 & 3538 && 7.826 [0.315] \\
1.204 & 7.439 [0.270] & 1.05 & 3588 && 7.440 [0.330] \\
1.209 & 7.596 [0.320] & 1.57 & 4345 && 7.515 [0.324] \\
1.213 & 7.735 [0.230] & 0.86 & 3171 && 7.522 [0.320] \\
1.218 & 8.105 [0.273] & 1.33 & 3953 && 8.027 [0.301] \\
1.223 & 7.596 [0.341] & 1.84 & 4617 && 7.282 [0.328] \\
1.227 & 7.905 [0.258] & 1.15 & 3637 && 7.846 [0.304] \\
1.232 & 8.027 [0.346] & 2.13 & 4951 && 7.942 [0.301] \\
1.236 & 7.686 [0.251] & 1.03 & 3442 && 7.521 [0.317] \\
1.241 & 7.424 [0.335] & 1.73 & 4434 && 6.958 [0.341] \\
1.246 & 7.828 [0.283] & 1.36 & 3946 && 7.773 [0.306] \\
1.250 & 7.950 [0.351] & 2.18 & 4969 && 7.573 [0.313] \\
1.255 & 8.140 [0.293] & 1.58 & 4247 && 8.111 [0.293] \\
1.260 & 7.542 [0.301] & 1.43 & 4046 && 7.412 [0.321] \\
1.264 & 7.561 [0.270] & 1.17 & 3644 && 7.293 [0.325] \\
1.269 & 7.215 [0.312] & 1.42 & 4002 && 6.974 [0.339] \\
1.274 & 7.912 [0.276] & 1.36 & 3883 && 7.657 [0.306] \\
1.278 & 8.033 [0.279] & 1.43 & 3982 && 7.998 [0.293] \\
1.283 & 8.141 [0.267] & 1.35 & 3867 && 8.126 [0.288] \\
1.288 & 7.550 [0.235] & 0.92 & 3159 && 7.418 [0.312] \\
1.292 & 7.464 [0.343] & 1.97 & 4553 && 7.149 [0.320] \\
1.297 & 8.013 [0.327] & 2.08 & 4657 && 7.789 [0.292] \\
1.301 & 8.059 [0.295] & 1.70 & 4226 && 7.745 [0.295] \\
1.306 & 8.206 [0.294] & 1.72 & 4278 && 8.002 [0.287] \\
1.311 & 7.435 [0.325] & 1.75 & 4288 && 7.735 [0.295] \\
1.315 & 7.955 [0.316] & 1.95 & 4461 && 7.617 [0.296] \\
1.320 & 7.844 [0.273] & 1.43 & 3804 && 7.717 [0.291] \\
1.325 & 8.461 [0.234] & 1.19 & 3520 && 8.475 [0.269] \\
1.329 & 8.012 [0.254] & 1.26 & 3627 && 7.911 [0.288] \\
1.334 & 7.955 [0.248] & 1.19 & 3521 && 7.937 [0.287] \\
1.339 & 7.273 [0.313] & 1.59 & 4063 && 7.054 [0.322] \\
1.343 & 8.083 [0.260] & 1.35 & 3752 && 7.930 [0.287] \\
1.348 & 8.193 [0.292] & 1.74 & 4273 && 8.201 [0.278] \\
1.353 & 8.189 [0.263] & 1.42 & 3851 && 8.034 [0.284] \\
1.357 & 8.571 [0.226] & 1.14 & 3455 && 8.661 [0.263] \\
1.362 & 8.306 [0.254] & 1.36 & 3761 && 7.983 [0.285] \\
1.366 & 7.876 [0.269] & 1.35 & 3779 && 7.784 [0.294] \\
1.371 & 7.241 [0.288] & 1.31 & 3733 && 7.019 [0.327] \\
1.376 & 8.271 [0.286] & 1.66 & 4224 && 8.211 [0.281] \\
1.380 & 7.755 [0.253] & 1.12 & 3497 && 7.625 [0.305] \\
1.385 & 7.185 [0.294] & 1.31 & 3775 && 7.063 [0.330] \\
1.390 & 7.631 [0.280] & 1.34 & 3825 && 7.524 [0.309] \\
1.394 & 7.902 [0.241] & 1.08 & 3403 && 7.787 [0.296] \\
1.399 & 7.898 [0.255] & 1.22 & 3597 && 7.701 [0.298] \\
1.404 & 7.616 [0.258] & 1.16 & 3518 && 7.523 [0.306] \\
1.408 & 8.306 [0.289] & 1.76 & 4297 && 8.458 [0.270] \\
1.413 & 7.857 [0.226] & 1.00 & 3185 && 7.499 [0.300] \\
1.418 & 6.891 [0.329] & 1.61 & 4059 && 6.563 [0.344] \\
1.422 & 8.018 [0.278] & 1.53 & 3989 && 8.076 [0.281] \\
1.427 & 7.342 [0.328] & 1.79 & 4303 && 7.377 [0.308] \\
1.432 & 8.046 [0.212] & 0.89 & 3053 && 7.991 [0.284] \\
1.436 & 7.819 [0.281] & 1.54 & 3937 && 7.621 [0.293] \\
1.441 & 8.028 [0.242] & 1.18 & 3483 && 7.813 [0.289] \\
1.445 & 8.427 [0.233] & 1.18 & 3517 && 8.607 [0.265] \\
1.450 & 7.419 [0.296] & 1.53 & 3929 && 7.430 [0.301] \\
1.455 & 7.692 [0.272] & 1.38 & 3744 && 7.379 [0.305] \\
1.459 & 8.261 [0.239] & 1.19 & 3535 && 8.253 [0.276] \\
1.464 & 8.052 [0.239] & 1.10 & 3446 && 7.835 [0.295] \\
1.469 & 7.318 [0.245] & 0.96 & 3218 && 7.002 [0.330] \\
1.473 & 7.988 [0.235] & 1.02 & 3362 && 8.080 [0.290] \\
1.478 & 8.339 [0.221] & 0.99 & 3305 && 8.179 [0.285] \\
1.483 & 7.779 [0.239] & 1.00 & 3326 && 7.730 [0.302] \\
1.487 & 7.916 [0.180] & 0.60 & 2544 && 7.791 [0.297] \\
1.492 & 7.709 [0.220] & 0.85 & 3034 && 7.379 [0.314] \\
1.497 & 8.431 [0.218] & 0.97 & 3282 && 8.651 [0.270] \\
1.501 & 7.801 [0.211] & 0.78 & 2940 && 8.027 [0.292] \\
1.506 & 7.541 [0.238] & 0.95 & 3209 && 7.250 [0.320] \\
1.510 & 7.906 [0.239] & 1.08 & 3380 && 7.829 [0.293] \\
1.515 & 7.506 [0.249] & 1.05 & 3336 && 7.280 [0.315] \\
1.520 & 7.624 [0.196] & 0.66 & 2672 && 7.594 [0.305] \\
1.524 & 7.865 [0.259] & 1.22 & 3632 && 7.819 [0.296] \\
1.529 & 7.707 [0.254] & 1.14 & 3493 && 7.751 [0.297] \\
1.534 & 7.780 [0.240] & 1.03 & 3337 && 7.776 [0.299] \\
1.538 & 7.305 [0.255] & 1.01 & 3331 && 7.141 [0.327] \\
1.543 & 7.378 [0.458] & 3.29 & 6086 && 7.251 [0.325] \\
1.548 & 7.773 [0.277] & 1.34 & 3853 && 7.649 [0.308] \\
1.552 & 7.739 [0.247] & 1.05 & 3422 && 7.627 [0.309] \\
1.557 & 8.317 [0.223] & 0.98 & 3315 && 8.152 [0.290] \\
1.562 & 7.987 [0.209] & 0.80 & 2985 && 7.679 [0.307] \\
1.566 & 7.515 [0.213] & 0.72 & 2860 && 7.354 [0.324] \\
1.571 & 8.324 [0.222] & 0.94 & 3281 && 8.541 [0.280] \\
1.575 & 8.242 [0.232] & 1.02 & 3400 && 8.219 [0.290] \\
1.580 & 7.739 [0.230] & 0.89 & 3165 && 7.480 [0.318] \\
1.585 & 7.902 [0.204] & 0.72 & 2860 && 7.927 [0.301] \\
1.589 & 7.823 [0.244] & 1.02 & 3391 && 7.625 [0.311] \\
1.594 & 7.863 [0.253] & 1.11 & 3531 && 7.542 [0.315] \\
1.599 & 8.002 [0.238] & 1.00 & 3381 && 7.799 [0.307] \\
1.603 & 7.578 [0.261] & 1.08 & 3510 && 7.537 [0.317] \\
1.608 & 7.894 [0.239] & 0.94 & 3340 && 7.963 [0.306] \\
1.613 & 8.069 [0.255] & 1.11 & 3654 && 8.137 [0.302] \\
1.617 & 8.044 [0.227] & 0.87 & 3248 && 8.141 [0.302] \\
1.622 & 7.980 [0.244] & 0.99 & 3459 && 7.830 [0.314] \\
1.627 & 8.002 [0.235] & 0.95 & 3342 && 7.851 [0.310] \\
1.631 & 7.912 [0.264] & 1.13 & 3698 && 8.306 [0.297] \\
1.636 & 7.878 [0.231] & 0.85 & 3224 && 7.648 [0.323] \\
1.640 & 8.008 [0.221] & 0.79 & 3140 && 8.025 [0.313] \\
1.645 & 7.879 [0.234] & 0.85 & 3260 && 7.850 [0.319] \\
\hline
\end{longtable}
\end{flushleft} 
}

\end{document}